\documentclass{layout}
\usepackage[dvipsnames]{xcolor}

\begin{document}

\markboth{K. Lichtenegger, W. Schappacher}
{Phase Transition in Stochastic Forest Fire Model; Effects of Definition of Neighbourhood}

\catchline{}{}{}{}{}
%%%%%%%%%%%%%%%%%%%%%%%%%%%%%%%%%%%%%%%%%%%%%%%%%%%%%%%%%%%%%%%%%%%%

\title{PHASE TRANSITION IN A STOCHASTIC FOREST FIRE MODEL
  AND EFFECTS OF THE DEFINITION OF NEIGHBOURHOOD}

\author{KLAUS LICHTENEGGER}

\address{Institute of Physics, Universit\"atsplatz 5,\\
Graz University, 8010 Graz, Austria\\
klaus.lichtenegger@uni-graz.at}

\author{WILHELM SCHAPPACHER}

\address{Institute of Mathematics and Scientific Computing, Heinrichstra§e 36 \\
Graz University, 8010 Graz, Austria\\
wilhelm.schappacher@uni-graz.at}

\maketitle

\begin{history}
\received{Day Month Year}
\revised{Day Month Year}
\end{history}

\begin{abstract}
We present results on a stochastic forest fire model,
where the influence of the neighbour trees is treated
in a more realistic way than usual and the definition
of neighbourhood can be tuned by an additional parameter.

This model exhibits a surprisingly sharp phase transition
which can be shifted by redefinition of neighbourhood.
The results can also be interpreted in terms of
disease-spreading and are quite unsettling from the
epidemologist's point of view, since variation of one crucial
parameter only by a few percent can result in the change
from endemic to epidemic behaviour.

\end{abstract}

\ccode{PACS Nos.: 64.60.ah, 89.75.-k, 89.75.Fb, 89.75.kd}

%%%%%%%%%%%%%%%%%%%%%%%%%%%%%%%%%%%%%%%%%
\section{Introduction}
\label{sec:introduction}

Cellular Automata (CA) are recognized as an important modelling
tool in many fields of science, including also biology and especially
ecology. The main characteristics of the CA approach are summarized
in three rules (see~[\refcite{vNThSelfReprod}]):

\begin{enumerate}
\item[(R1)] Space and time are discrete: There is a (often one- or
  two-dimensional) grid of cells that is only viewed at distinct
  (equidistant) timesteps.
\item[(R2)] At each timestep, each cell has one (and only one) state taken
  from a limited number of possible states.
\item[(R3)] There are simple but universal deterministic update rules: The
  state of a cell at a given timestep depends only on its own
  state and the cell states in its neighbourhood, all taken at the
  previous step. These rules are the same for all cells and timesteps.
\end{enumerate}

A more detailed discussion of the CA approach and its history
can be found for example in~[\refcite{Lichtenegger:2005DA}]. CA models
which strictly respect these three rules (and, in many cases, make use of only two
cell states) have been intensively studied and sometimes can
give considerable insight in the foundations of complex behaviour.

Each rule can be modified in order to extend the range of systems which
can be simulated. We soften rule $(R3)$ by making the update rules still
universal, but stochastic, employing (pseudo)random numbers to determine
cell evolution.

A topic that cellular models often excel in is ecological modelling,
especially the study of disease spreading. In fact, there is a wide
variety of systems that can be modeled with CA, and disease spreading
models of this type have been successfully applied to plants (e.g.~[\refcite{Citrus}]),
animals (e.g.~[\refcite{MorleyFM}]) and, in part, human beings (e.g.~[\refcite{Avian}]).

Often CA models are not applied to special (realistic) cases, but general
properties are studied, as for example in~[\refcite{HorVertSpread}]
or~[\refcite{FlexAutDis}]. One of the most interesting (and most dramatic)
features of CA models is the possibility of ``phase transition'' (see for
example~[\refcite{clar-1997-55}, \refcite{CataDaisy}]).

In the following we will study such transitions as well -- where the
term \emph{phase transition} is used in a more general meaning than the one
common in physics since there does not necessarily exist any quantity 
analogous to free energy. Still, some order parameter can be found which
allows to clearly distinguish distinct phases of the system and on the
border there are indications for critical behaviour.
%Actually, for many models one finds \emph{self-organized criticality},
% to be discussed in subsection~\ref{ssec:OtherModels}.

%%%%%%%%%%%%%%%%%%%%%%%%%%%%%%%%%%%%%%%%%
\section{The Definition of  Neighbourhood}
\label{sec:defneigh}

On a 2D lattice, the neighbourhood of a cell may be defined
in many ways; the most popular ones are, however:
\begin{itemize}
\item \emph{von Neumann neighbourhood}: The neighbourhood consists
only of the four adjacent cells.
\item \emph{Moore neighbourhood}: In addition to the von Neumann
neighbourhood, also the four next-nearest-neighbours are taken into account.
\item \emph{extended Moore neighbourhood}: An even larger number
of cells is considered as neigbourhood.
\end{itemize}

\begin{figure}
  \begin{center}
    \includegraphics[width=2.8cm]{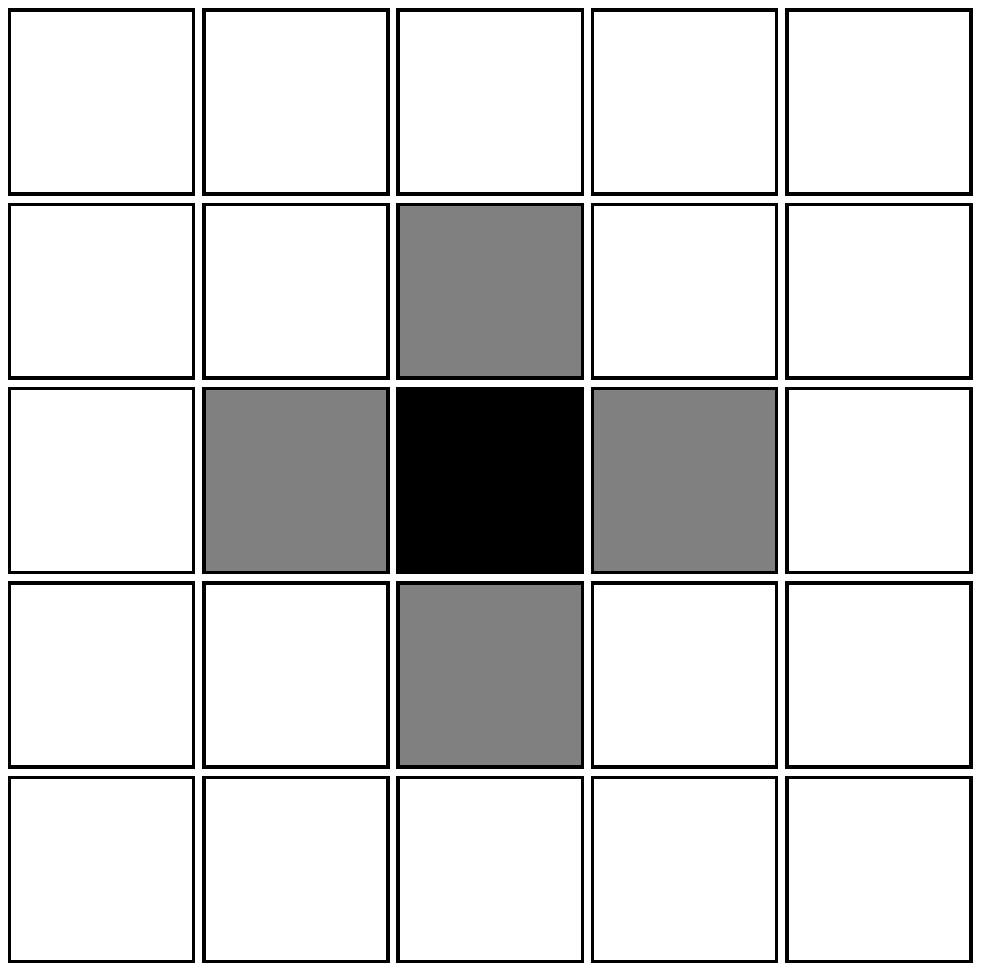} \put(-90,66){(a)} \hspace{3mm}
    \includegraphics[width=2.8cm]{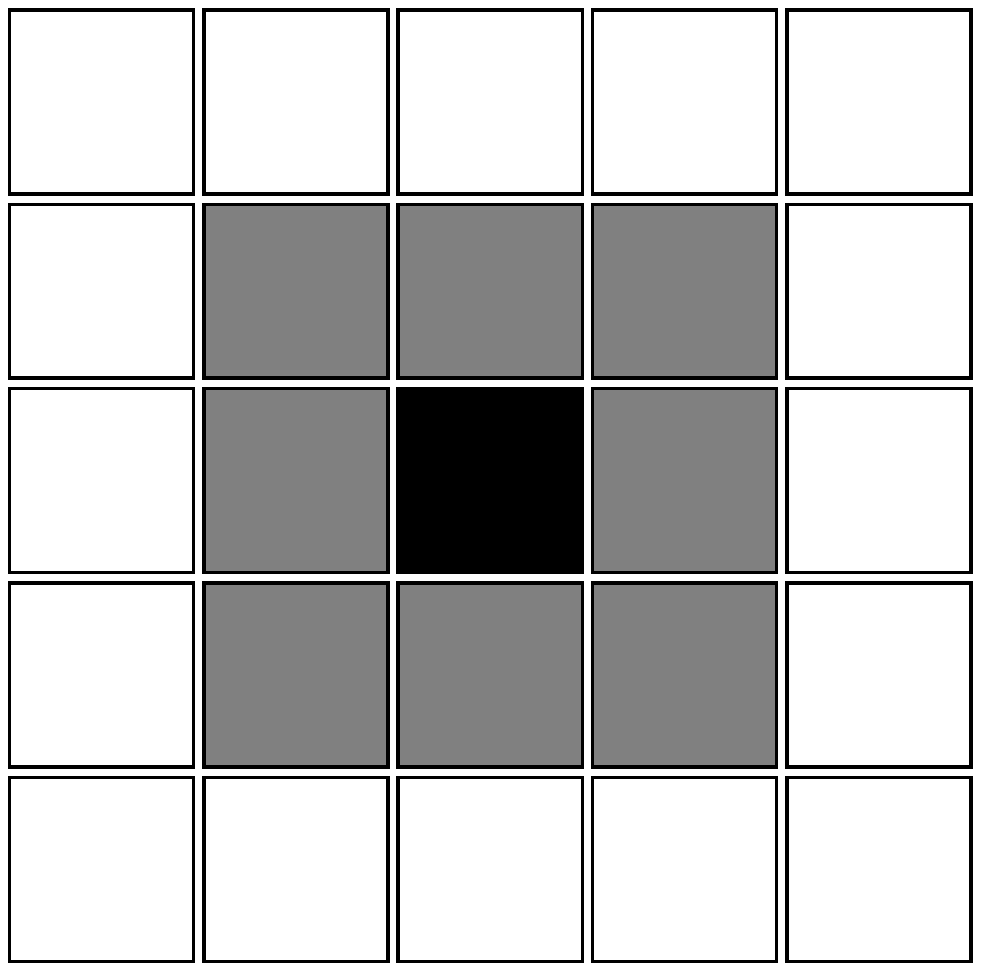} \put(-90,66){(b)} \hspace{3mm}
    \includegraphics[width=2.8cm]{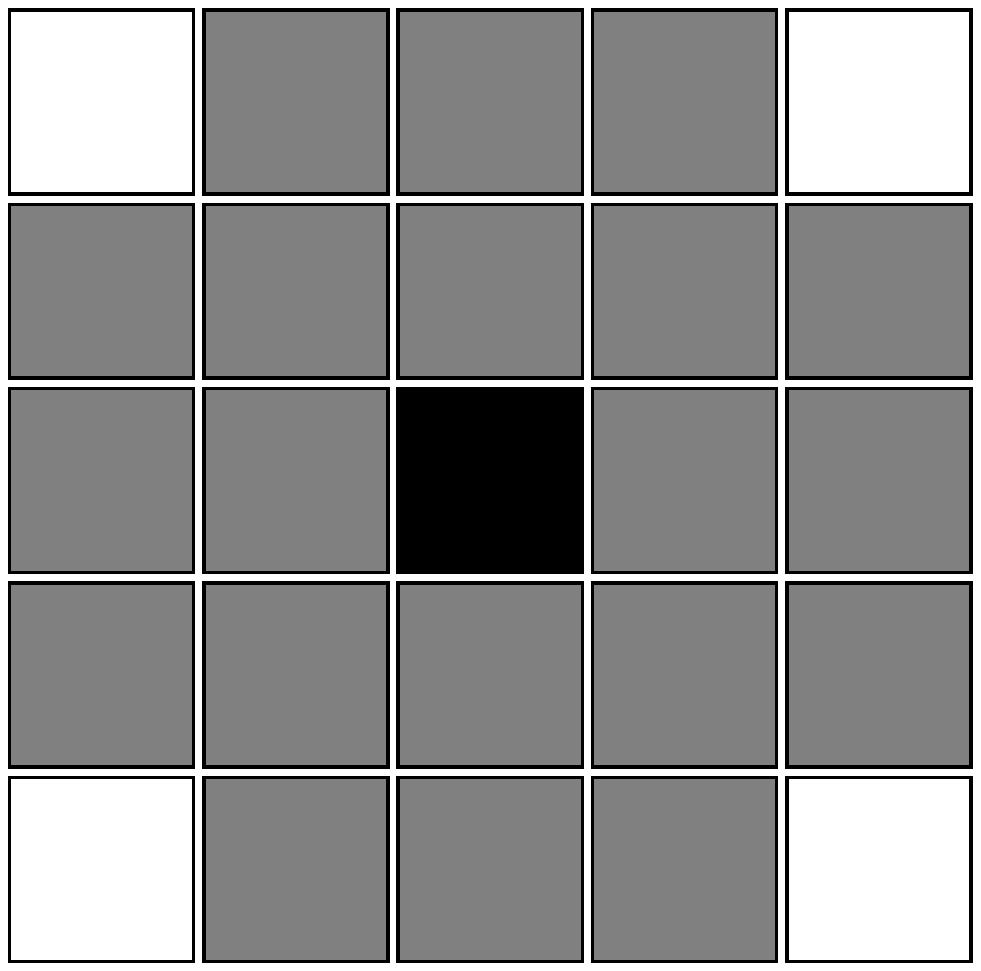} \put(-90,66){(c)} \hspace{3mm}
    \includegraphics[width=2.8cm]{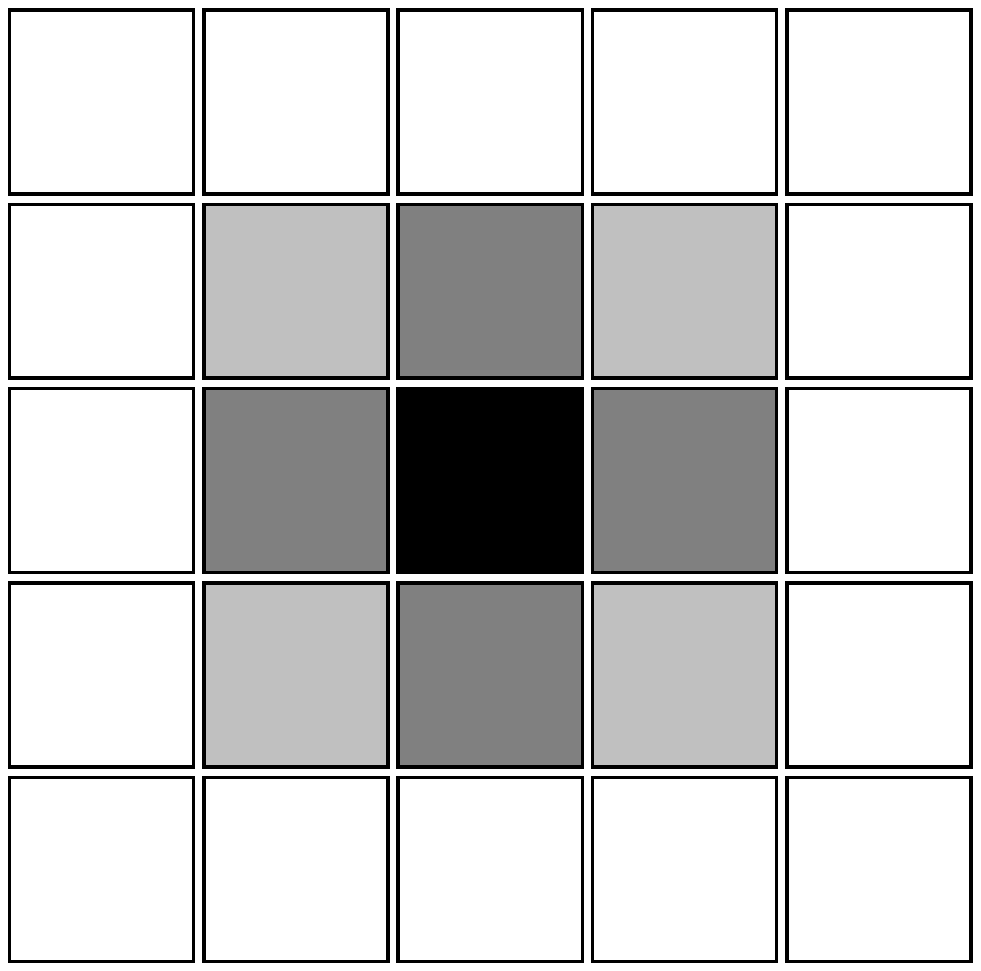} \put(-90,66){(d)}
  \end{center}
  \caption{Different neighbourhoods
     for Cellular Automata: (a) \emph{von Neumann} neighbourhood,
     (b) \emph{Moore} neighbourhood, (c) \emph{extended Moore}
     neighbourhood and (d) neighbourhood mediated by the
     Moore parameter $\pi_M$}
  \label{fig:ca_neighbours}
\end{figure}

\noindent These types of neighbourhood are illustrated in
figure~\ref{fig:ca_neighbours}.
The neighbourhood type and the number of cell states
determine the number of rules one has to specify (in principle).
For a neighbourhood of $N$ cells
and $s$ possible cell states one needs $s^{N+1}$ rules.
(This means, for example, $2^9=512$ for a two-state
model on a 2D grid with Moore neighbourhood as for example
Conway's Game of Life.)

Therefore in many models, the updates rules
depend only on the number of neigbours in a certain
state, but not on their actual position. This of course
greatly simplifies the speficication of update rules.

When doing this, an interesting type of rules are those that use an
extended neighbourhood (at least Moore-type), but reduce
the importance of non-adjacent cells for the update.
For update rules that only depend on the number of
neigbours in a certain state, their number $N$ can be
redefined as
\begin{equation}
  N:=N_{\mathrm{adj}}+\pi_M\cdot N_{\mathrm{non-adj}}
\end{equation}
where $N_{\mathrm{adj}}$ denotes the number of adjacent
and $N_{\mathrm{non-adj}}$ the number of non-adjacent
cells that are nevertheless regarded as neighbours.
In case of the next-nearest-neighbours, the \emph{Moore parameter}
$\pi_M$ mediates the transition from a pure von Neumann
neighbourhood ($\pi_M=0$) to a full Moore neighbourhood
($\pi_M=1$).
The accordingly modified neighbourhood is
also sketched in figure \ref{fig:ca_neighbours}.
Models of this kind have been studied in this paper, and the
influence of $\pi_M$ on certain results has been checked.

%%%%%%%%%%%%%%%%%%%%%%%%%%%%%%%%%%%%%%%%%
\section{The Forest Fire Model}
\label{sec:ForFirModel}

Our model is a variation of wide-spread forest fire models
(see~[\refcite{BCT:1990}, \refcite{CBJ:1990}, \refcite{clar-1994},
\refcite{honecker-1996-229}, \refcite{clar-1996-8}, \refcite{XToys96},
\refcite{grassberger-2002}], which can also be re-interpreted as a
disease-spreading model.
In this article, however, the vocabulary will be mostly the one of
living, burning and burnt-out trees.

\subsection{Definition of the Model}
\label{ssec:ModelDef}

The forest fire simulation takes place on a quadratic two-dimensional grid
(in all the following simulations $128\times 128$) with periodic boundary
conditions and a mediated Moore neighbourhood as described in
section~\ref{sec:defneigh}. The states were encoded the following
way:

\begin{center}\begin{tabular}{c|l|l}
$x$ & colour & state of tree \\ \hline
$-1$ & black & dead (burnt-out) tree \\
$0$ & {\color{gray}white} & empty cell (or fire-resistant tree) \\
$1$ & {\color{ForestGreen}green} & living tree \\
$2$ & {\color{red}red} & burning tree
\end{tabular}\end{center}

\noindent The only possible transitions are those from
living to burning state, $[1\to 2]$, and those from
the burning state to the burnt-out one, $[2\to (-1)]$.
The probability that a living tree at $(i,\,j)$
catches fire is given by:
\begin{equation}\label{eqn:forfirp12}
p_{[1\to2]}\left(i,\,j,\,t\,|\;\mathcal{N}_{t-1}\right) =
\delta_{x_{i,\,j},\,1}\cdot \left\{
1-(1-p_b)^{N_{b,t-1}} \right\}
\end{equation}
where $p_b$ denotes a fixed burning probability.
The state of the cell's neighbourhood at the
previous timestep is symbolized by the condition
complex $\mathcal{N}_{t-1}$, it explicitly enters
the formula as $N_{b,t-1}$,
the number of burning neighbours at $(t-1)$,
\begin{equation}\label{eqn:forfirNNlong}
N_{b,\,t-1} =
\left[\sum_{\begin{array}{c}\scriptstyle\Delta_{i,j}\in\{-1,0,1\} \\
\scriptstyle|\Delta_i|+|\Delta_j|=1 \end{array}}
\hspace{-6mm}\delta_{x_{i+\Delta_i,\,j+\Delta_j},2}
  \quad+\quad \pi_M\cdot \hspace{-5mm}
\sum_{\begin{array}{c}\scriptstyle\Delta_{i,j}\in\{-1,0,1\} \\
\scriptstyle|\Delta_i|+|\Delta_j|=2 \end{array}}\hspace{-6mm}
\delta_{x_{i+\Delta_j,\,j+\Delta_j},2}\right]_{t-1}\hspace{-5mm}.
\end{equation}
A burning tree is converted to a burnt-out one after one timestep, formally
\begin{equation}\label{eqn:forfirp2m1}
p_{[2\to-1]}\left(i,\,j,\,t\,|\;\mathcal{N}_{t-1}\right) = 
\delta_{x_{i,\,j},\,2}
\end{equation}

\begin{figure}
  \begin{center}
    \includegraphics[height=5cm]{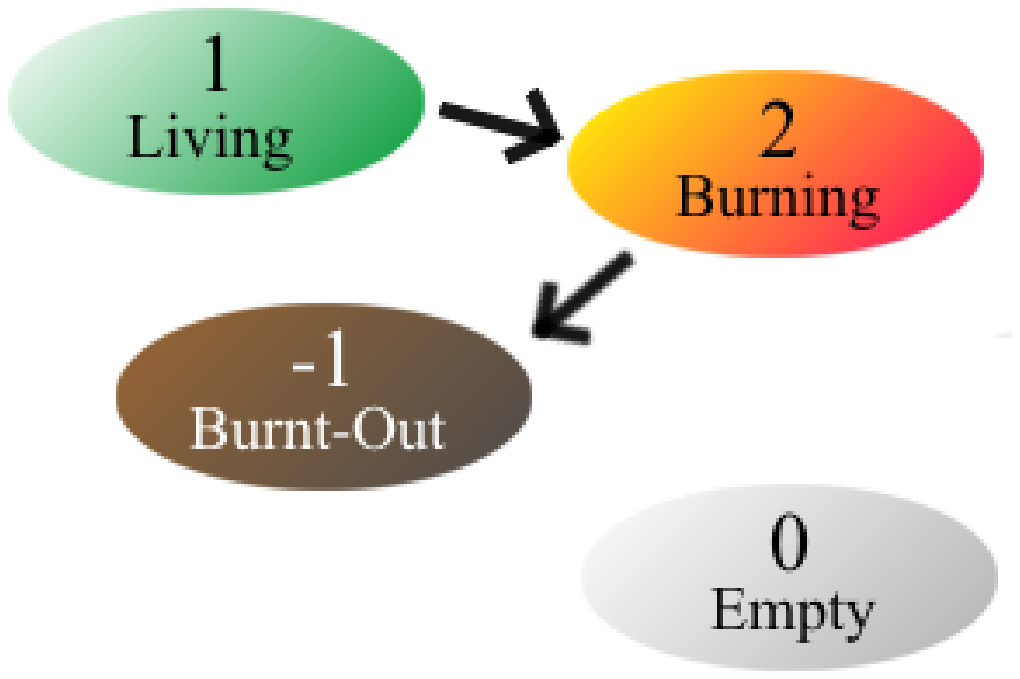}\put(-202,132){(a)}
    \hspace{1cm}
    \includegraphics[height=5cm]{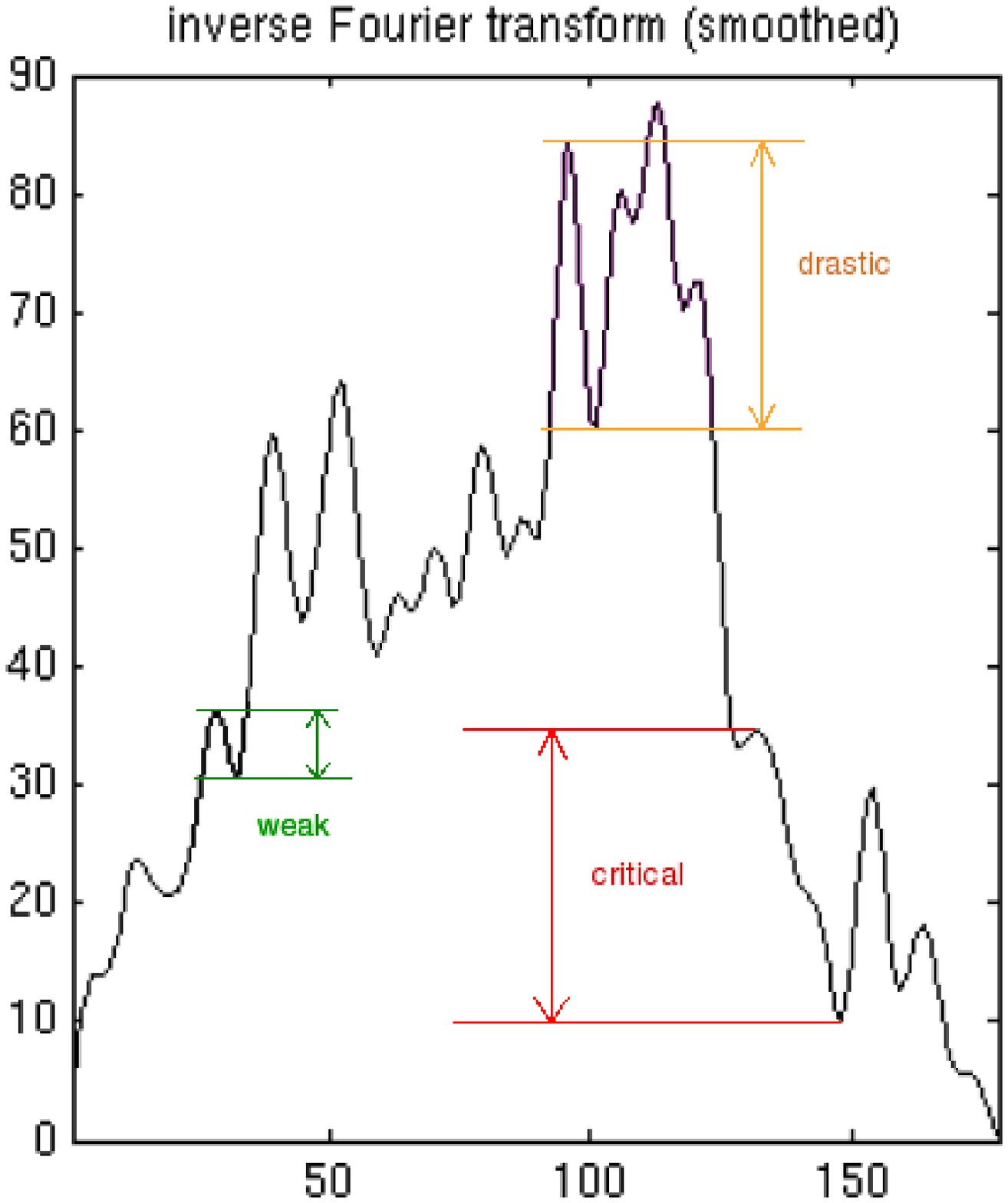}\put(-116,138){(b)}
  \end{center}
  \caption{(a) Sketch of the basic forest fire model
    (b) Expamples for weak, drastic and critical decreases
       of the number of burning trees in a smoothed time series}
  \label{fig:forfirsketch}
\end{figure}

A symbolic sketch of the model is given in figure~\ref{fig:forfirsketch}.a.
Equation \eqref{eqn:forfirp12} together with the definition
of $N_{b,\,t-1}$ expresses the increased
probability of a tree catching fire if it has more
than one burning neighbour. In the case $N_{b,\,t-1}=1$
the probability reduces to
\begin{equation}\label{eqn:forfirp12simple}
p_{[1\to2]}\left(x_{i,\,j}, t\,|\,\mathcal{N}_{t-1}\right) =
\delta_{x_{i,\,j},\,1}\cdot \left\{
1-(1-p_b) \right\} = \delta_{x_{i,\,j},\,1}\cdot p_b
\end{equation}

\subsection{A Typical Simulation Run}
\label{ssec:SimRun}

At the beginning of the simulation, living trees are placed
with a density $d_{\mathrm{start}}\in[0,1]$ while the
remaining cells stay empty. Than a few number of burning
trees are inserted -- starting with only one burning tree
would dramatically increase the probability of a
"{}false start"{}.

\begin{figure}
  \begin{center}
    \includegraphics[width=4.00cm]{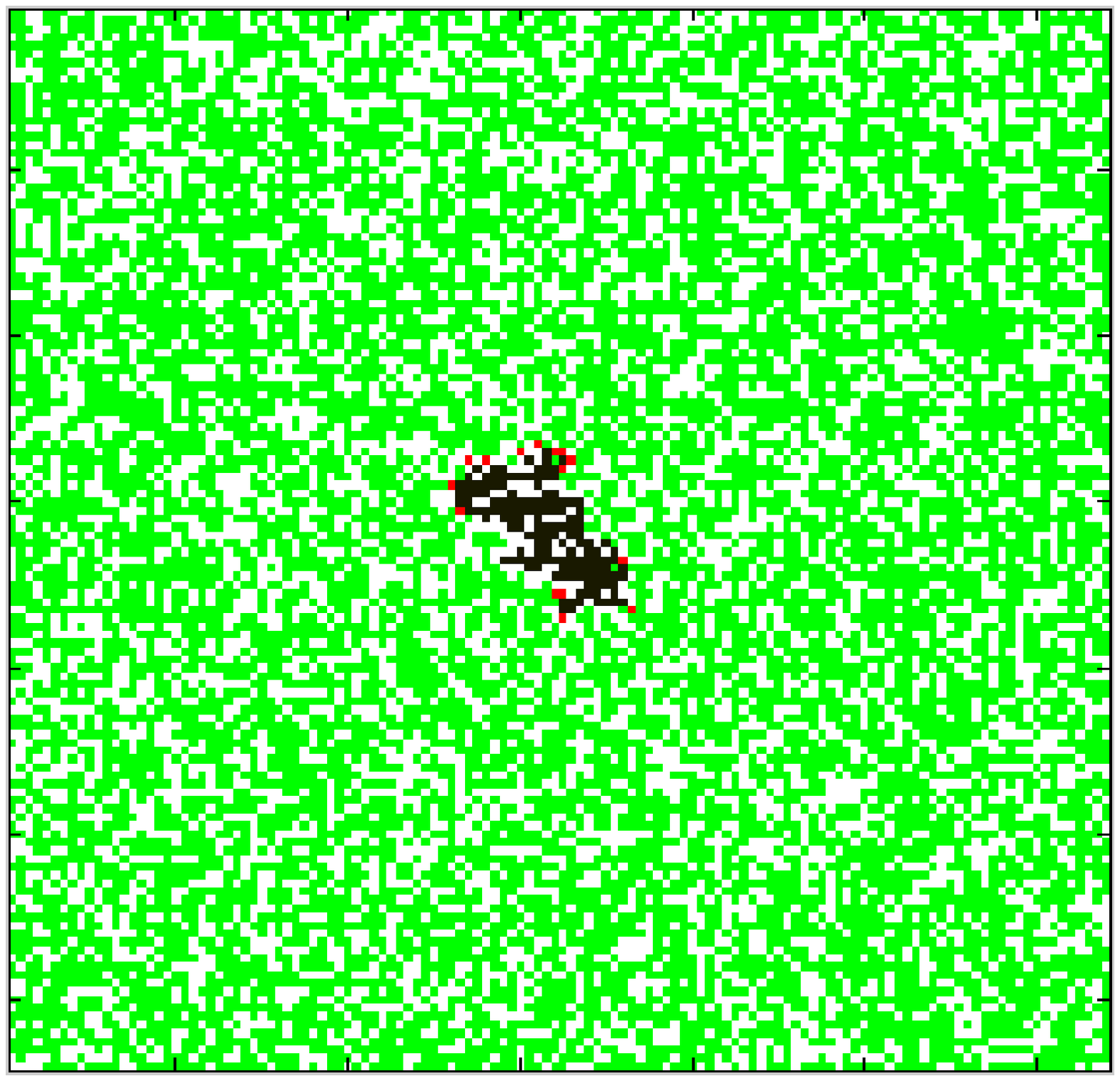} \put(-112,99){\bf\color{BrickRed}(a)} \hspace{1mm}
    \includegraphics[width=4.00cm]{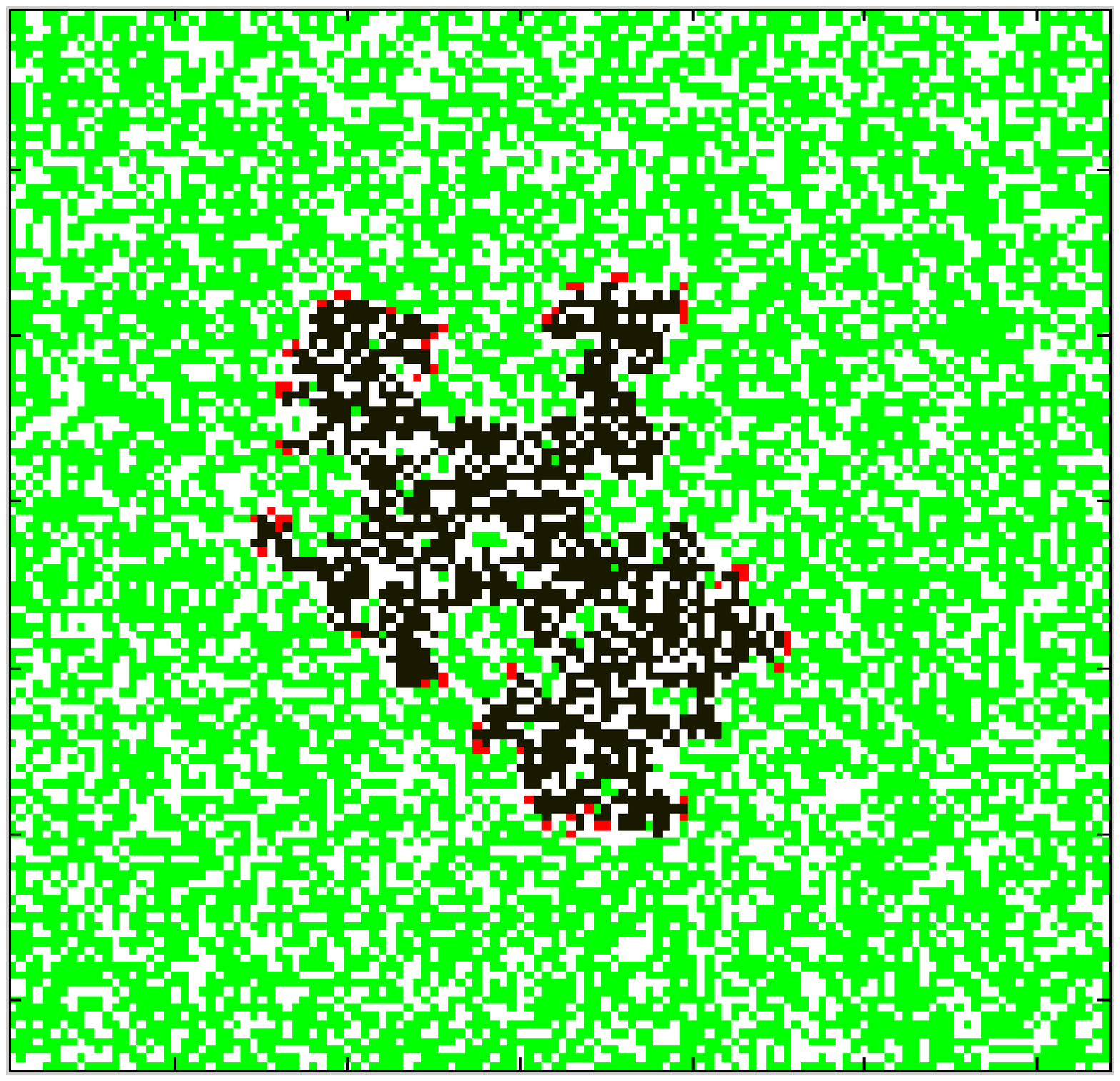} \put(-112,99){\bf\color{BrickRed}(b)} \hspace{1mm}
    \includegraphics[width=4.00cm]{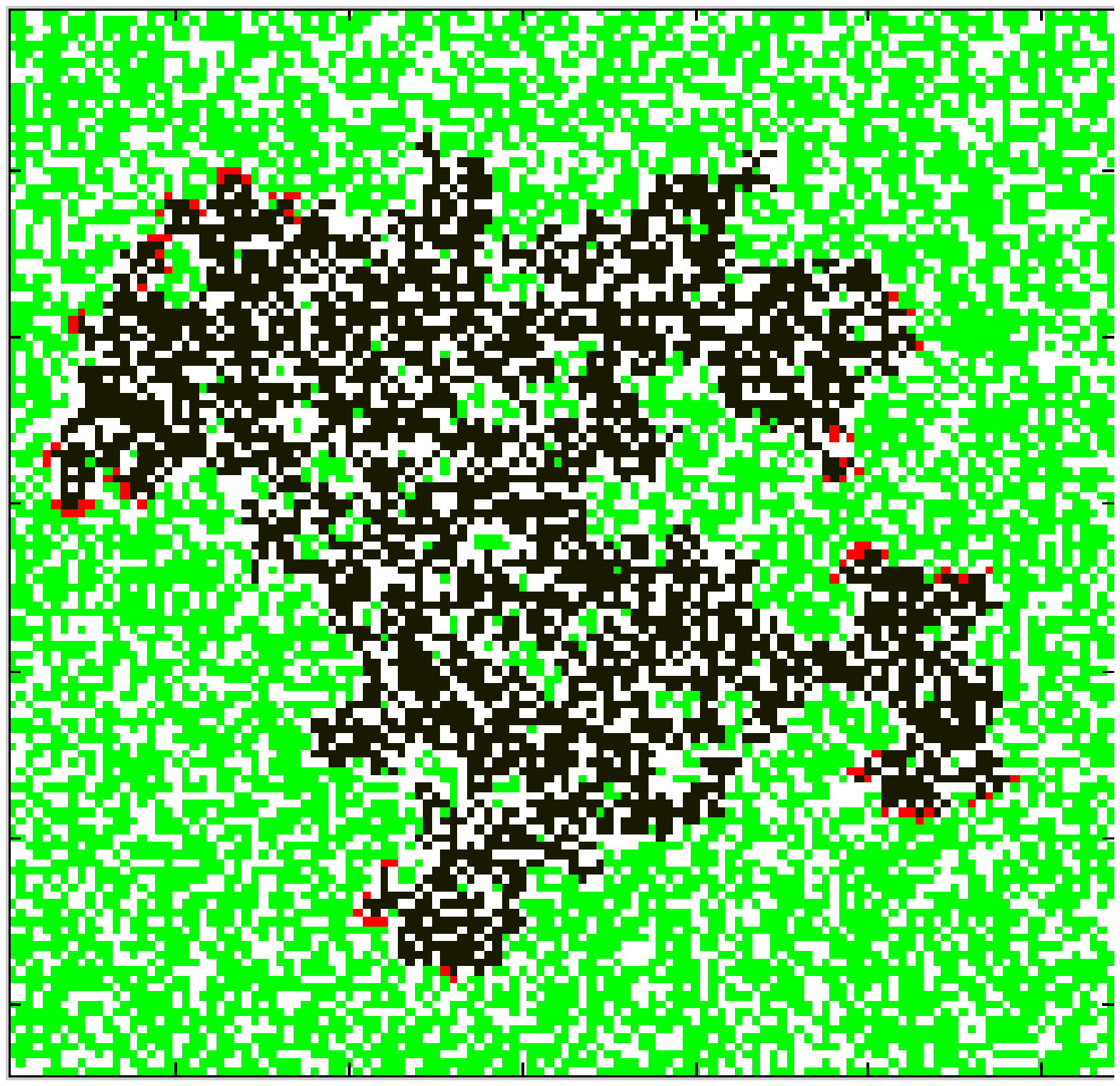} \put(-112,99){\bf\color{BrickRed}(c)} \\[6pt]
    \includegraphics[width=4.00cm]{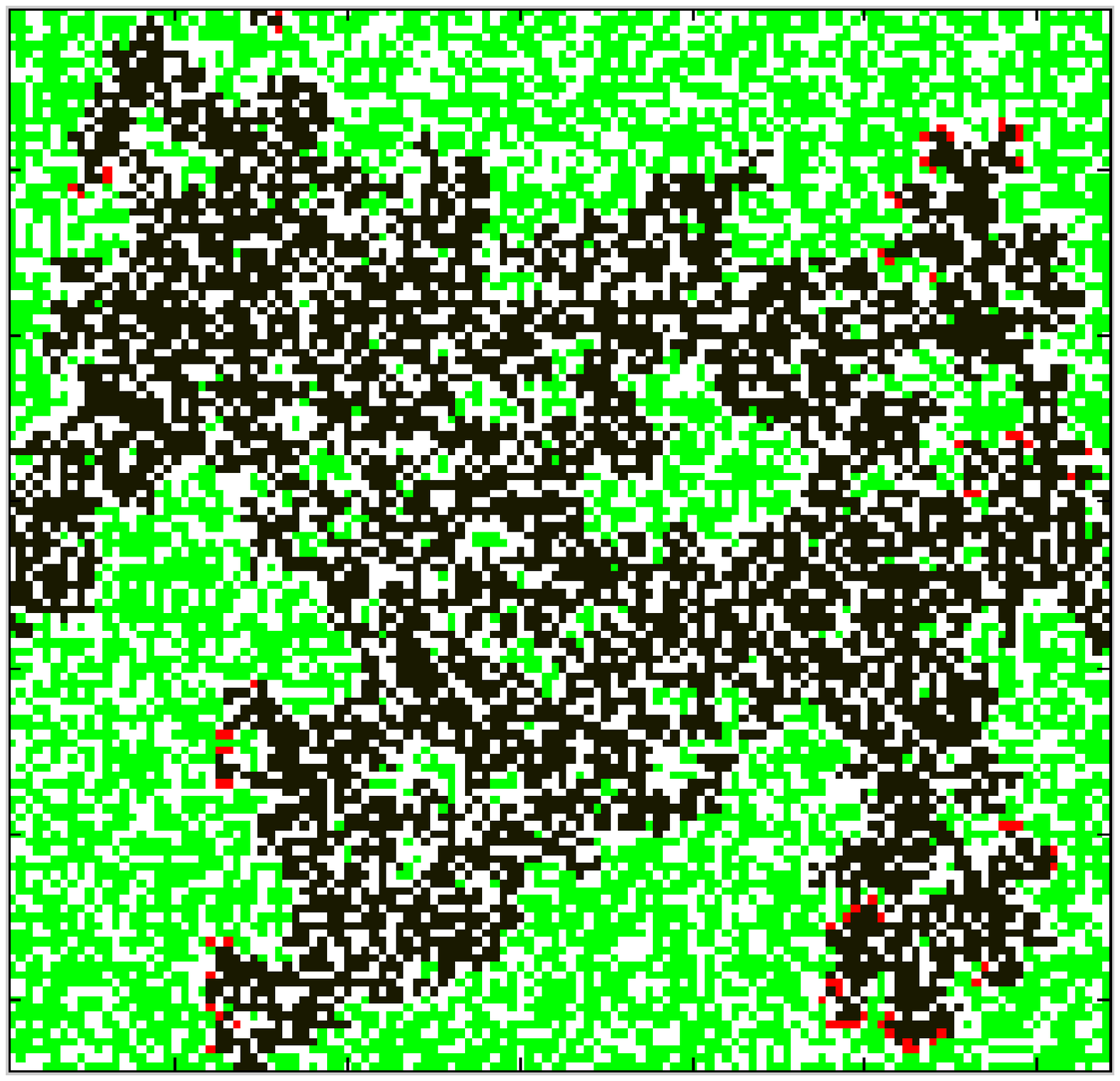} \put(-112,99){\bf\color{Red}(d)} \hspace{1mm}
    \includegraphics[width=4.00cm]{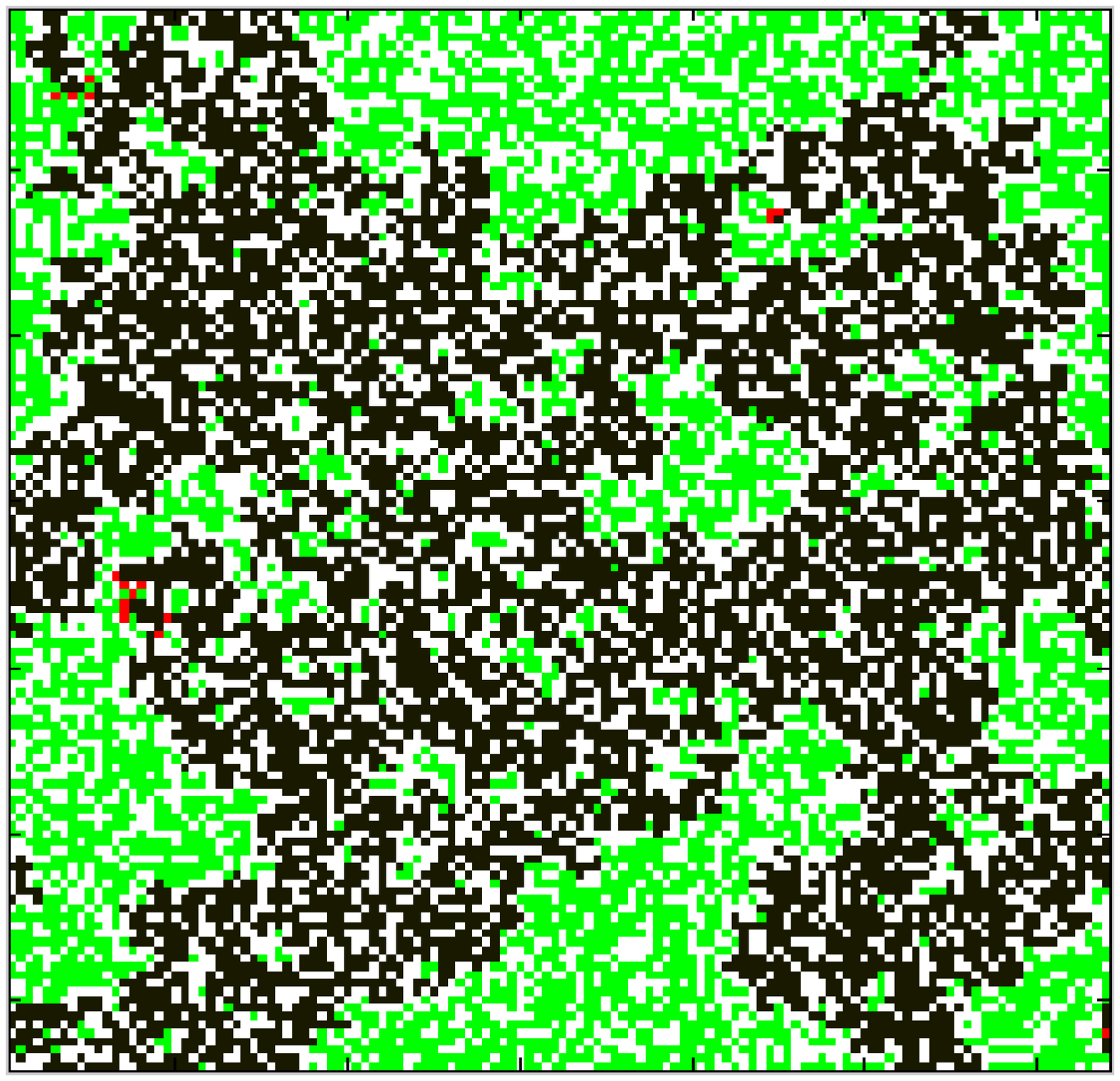} \put(-112,99){\bf\color{Red}(e)} \hspace{1mm}
    \includegraphics[width=4.00cm]{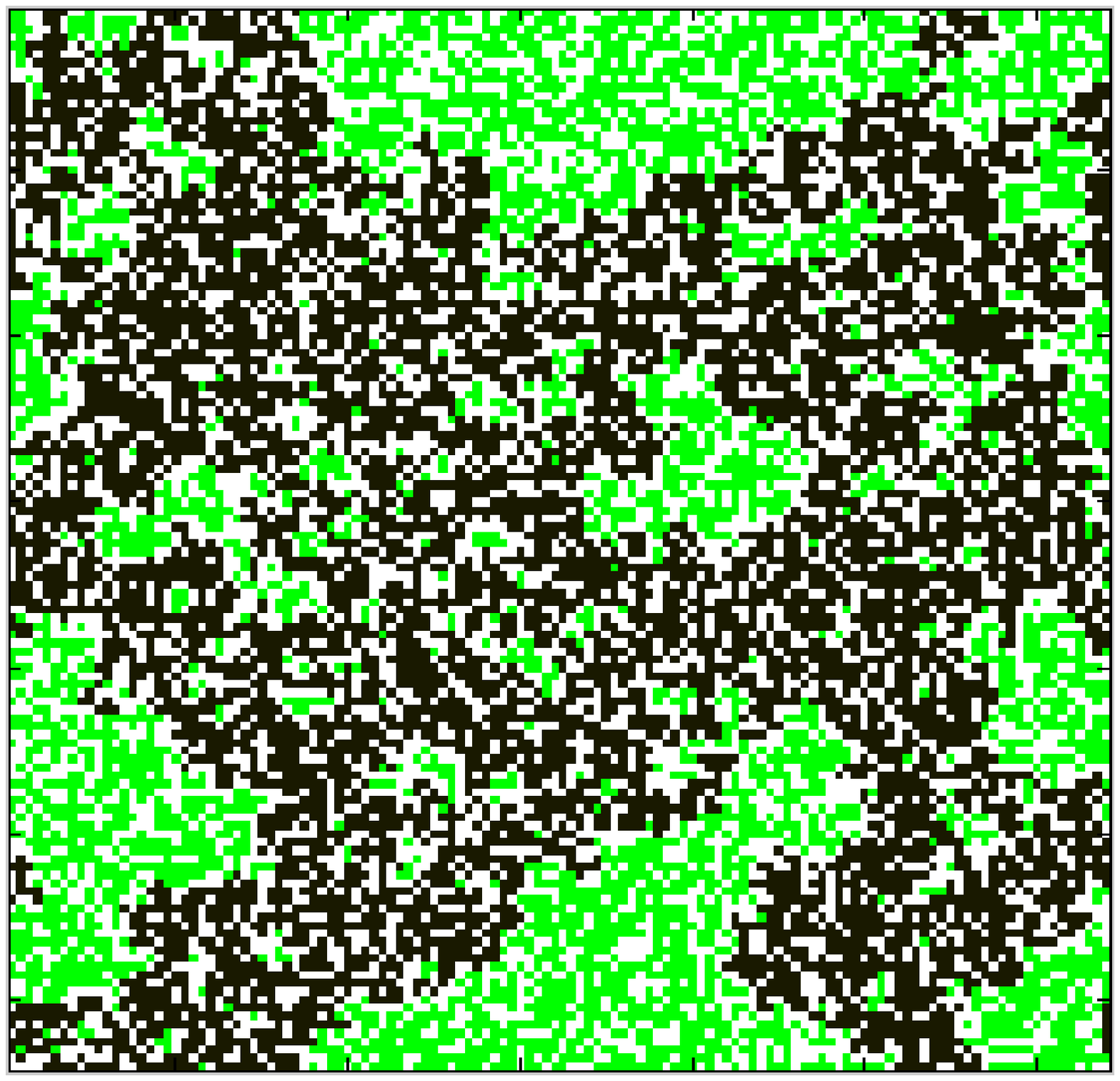} \put(-112,99){\bf\color{Red}(f)} \\[3pt]
    \includegraphics[width=12.5cm]{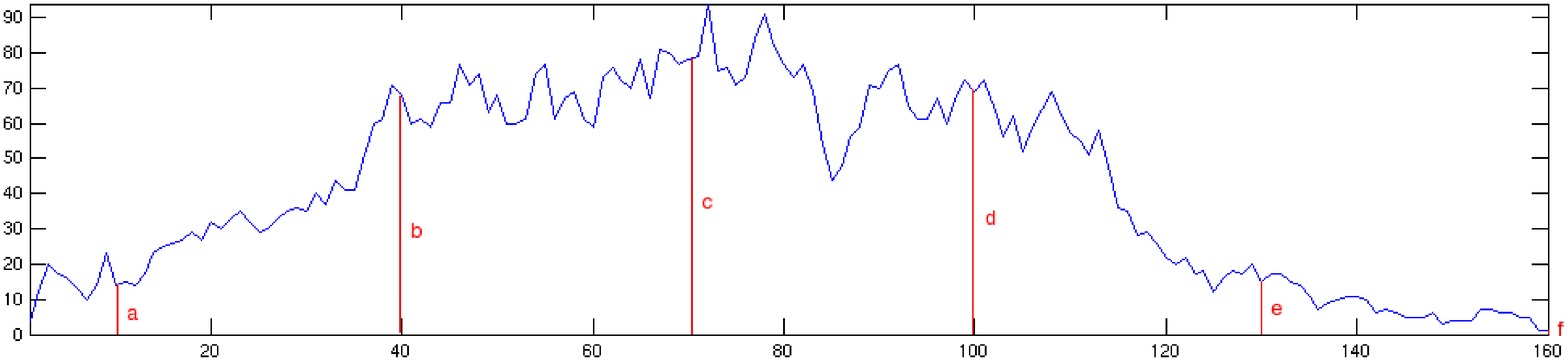}
  \end{center}
  \caption[Forest fire simulation for the critical case]
        {Snapshots of a forest fire simulation and
  	corresponding time series of burning trees
        (critical case, $p_b=d_{\mathrm{start}}=0.63$)}
  \label{fig:forfirerun}
\end{figure}

An example of a typical simulation run is shown in figure
\ref{fig:forfirerun} for the choice $p_b=d_{\mathrm{start}}=0.63$.
Even some runs with varied parameters already indicate
that there is a relatively sharp ``phase transition"
between those configurations where the fire dies out within
a few steps and those where almost the whole forest is
affected.

\begin{figure}
  \begin{center}
    \begin{tabular}{ccc}
     (a) & (b) & (c) \\[6pt]
      \includegraphics[width=3.7cm]{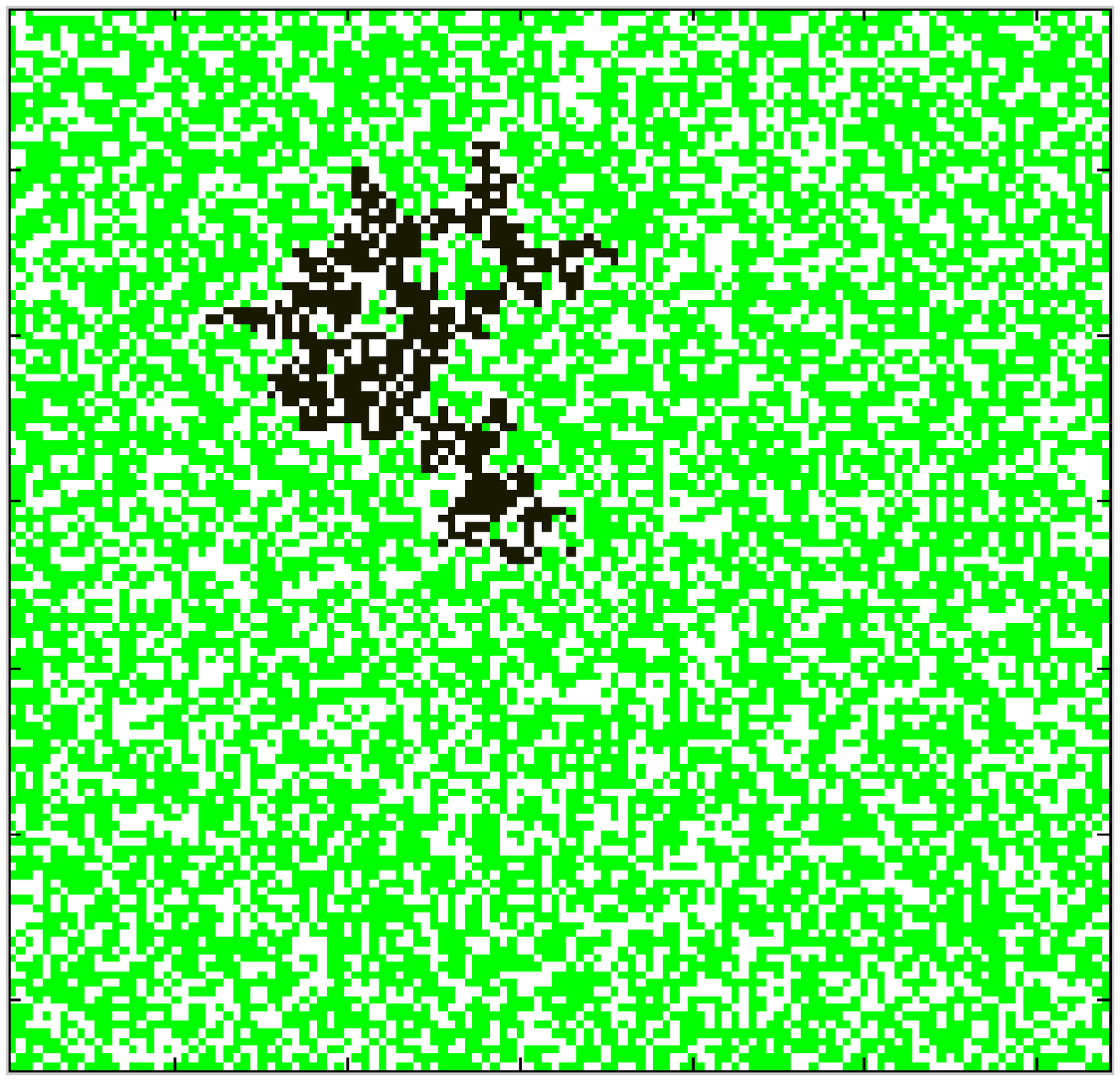} &
      \includegraphics[width=3.7cm]{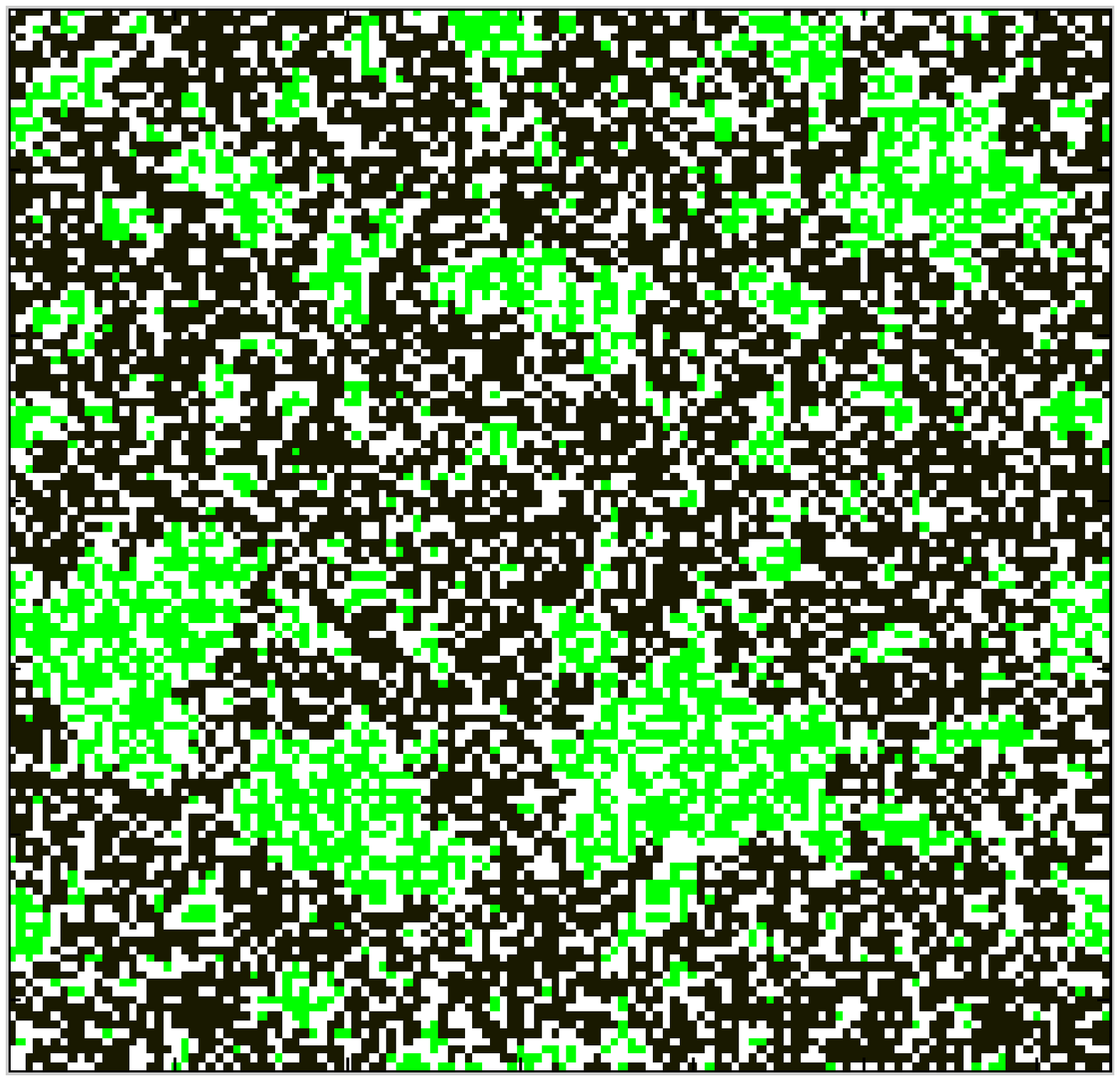} &
      \includegraphics[width=3.7cm]{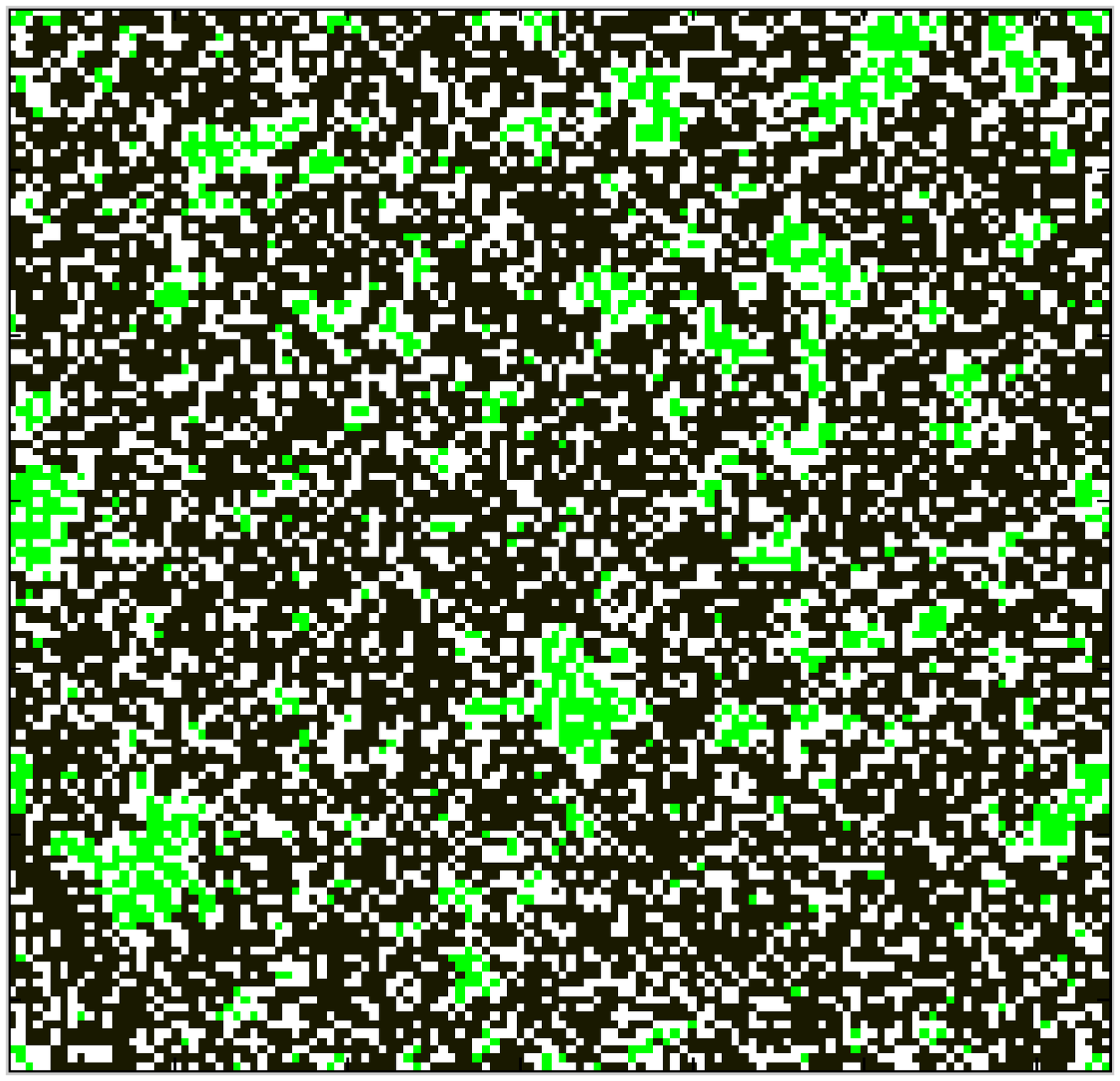} \\
      \includegraphics[height=3.7cm]{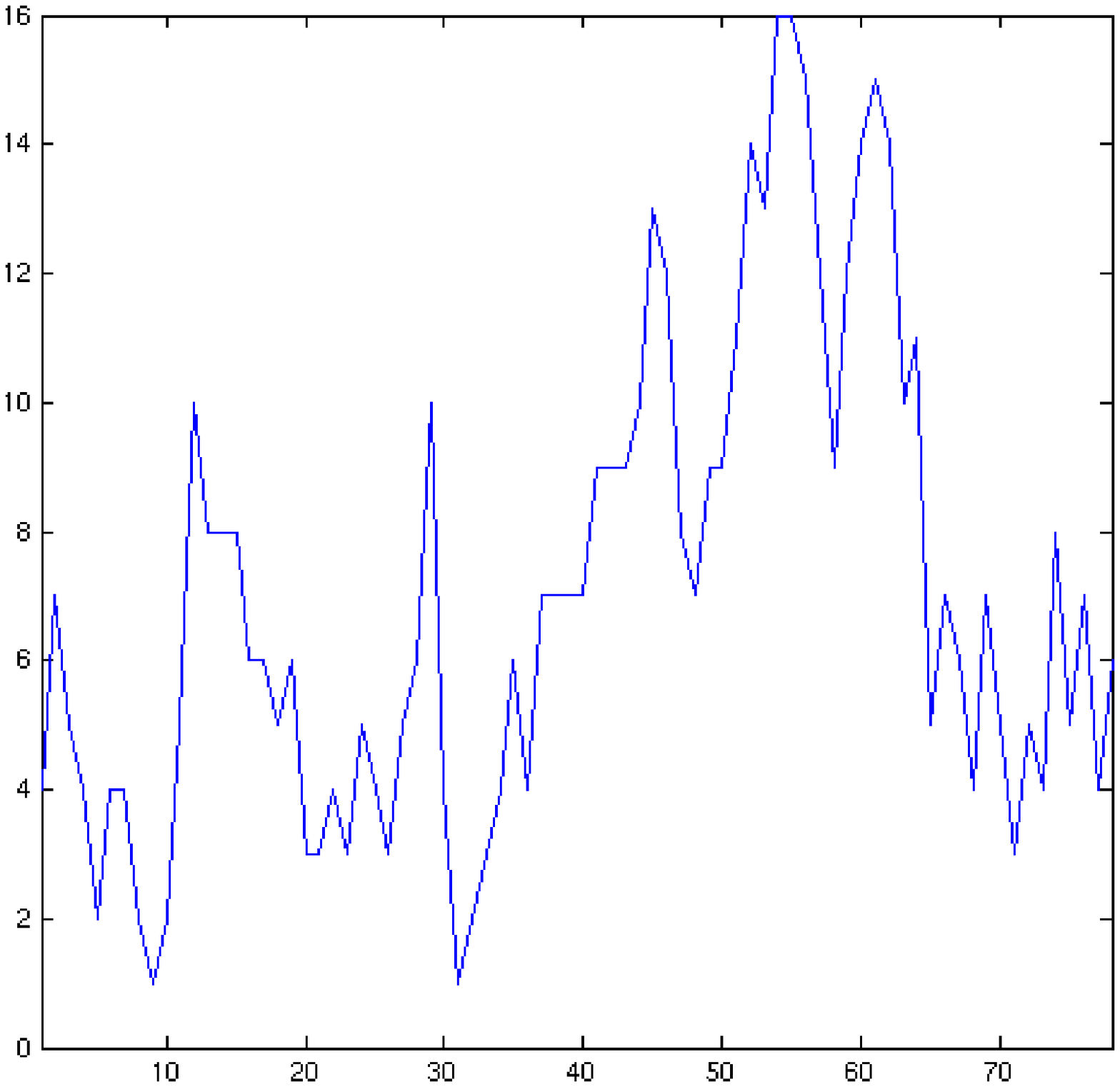} &
      \includegraphics[height=3.7cm]{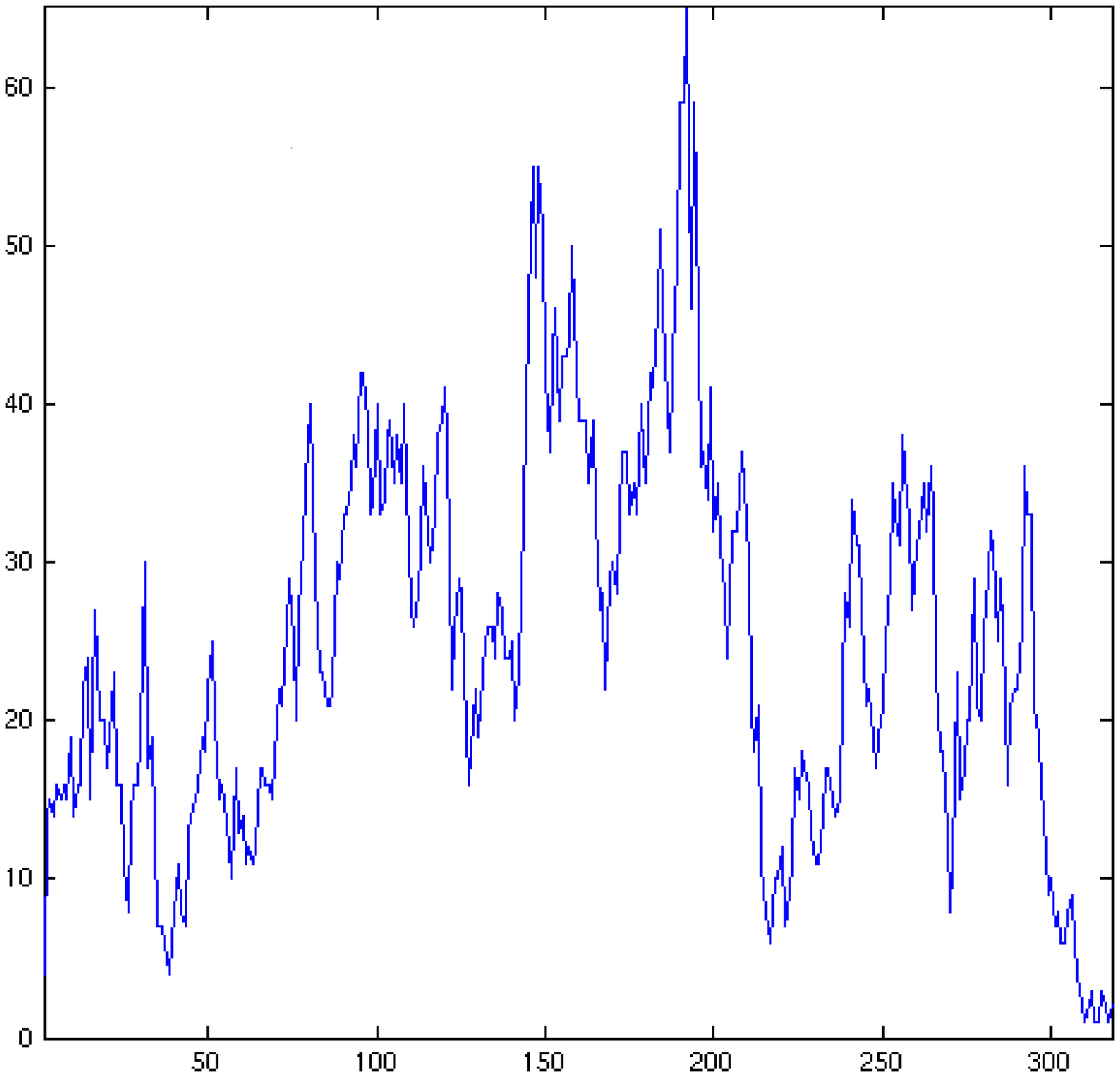} &
      \includegraphics[height=3.7cm]{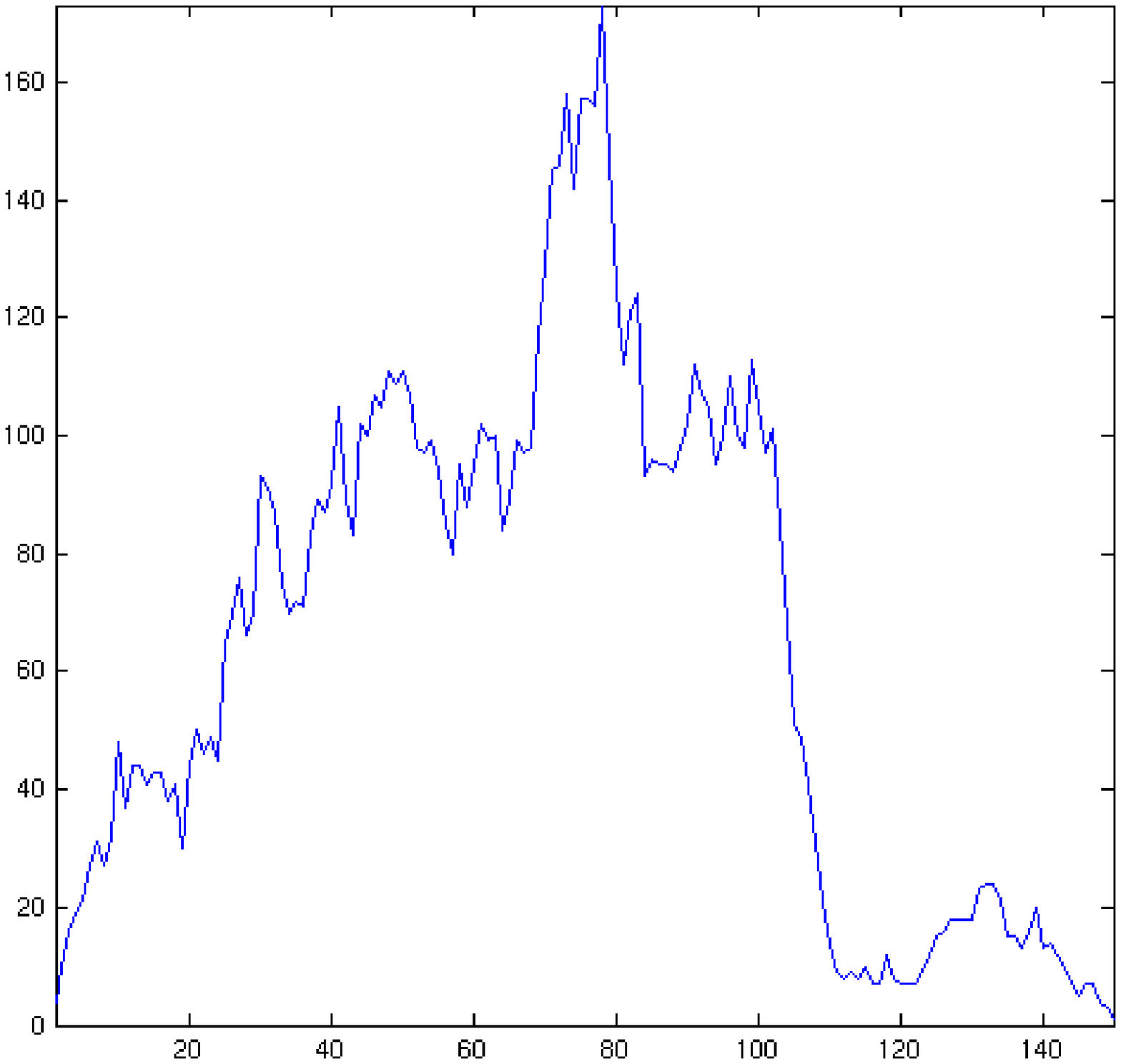}
    \end{tabular}
  \end{center}
  \caption{Typical end configuration
        and time series of burning trees in forest fire simulations:
	(a) subcritical case ($p_b=d_{\mathrm{start}}=0.60$),
    	(b) critical case ($p_b=d_{\mathrm{start}}=0.63$) and
    	(c) supercritical case ($p_b=d_{\mathrm{start}}=0.66$).
	Note the different timescales, in particular the fact that in
	case (b) the fire lasts about twice as long as in case (c).}
  \label{fig:forfireend}
\end{figure}

\begin{figure}
  \begin{center}
    \includegraphics[height=3.2cm]{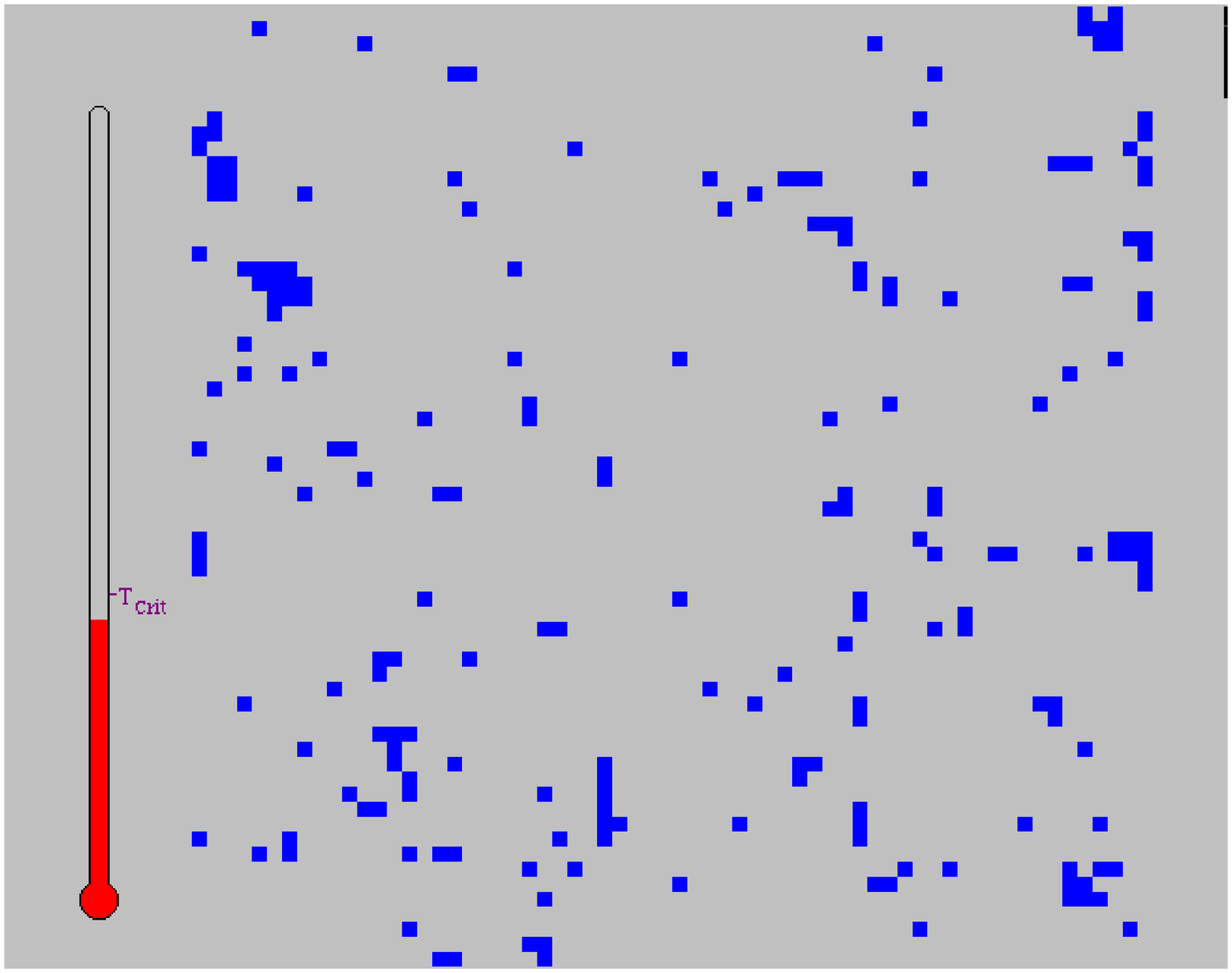} \put(-115,86){\tiny $T\!\!<\!\!T_c$} \hspace{1mm}
    \includegraphics[height=3.2cm]{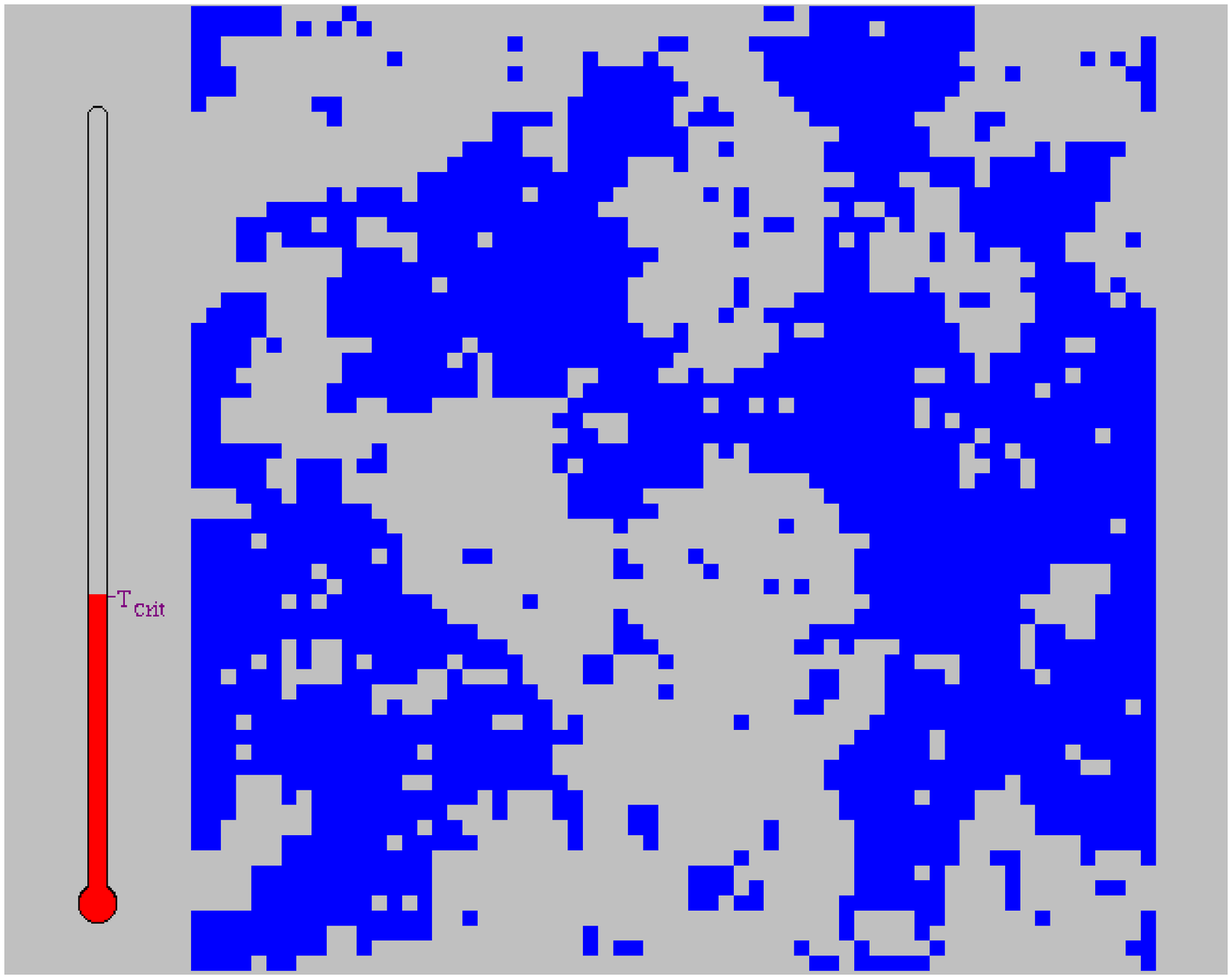} \put(-115,86){\tiny $T\!\!=\!\!T_c$} \hspace{1mm}
    \includegraphics[height=3.2cm]{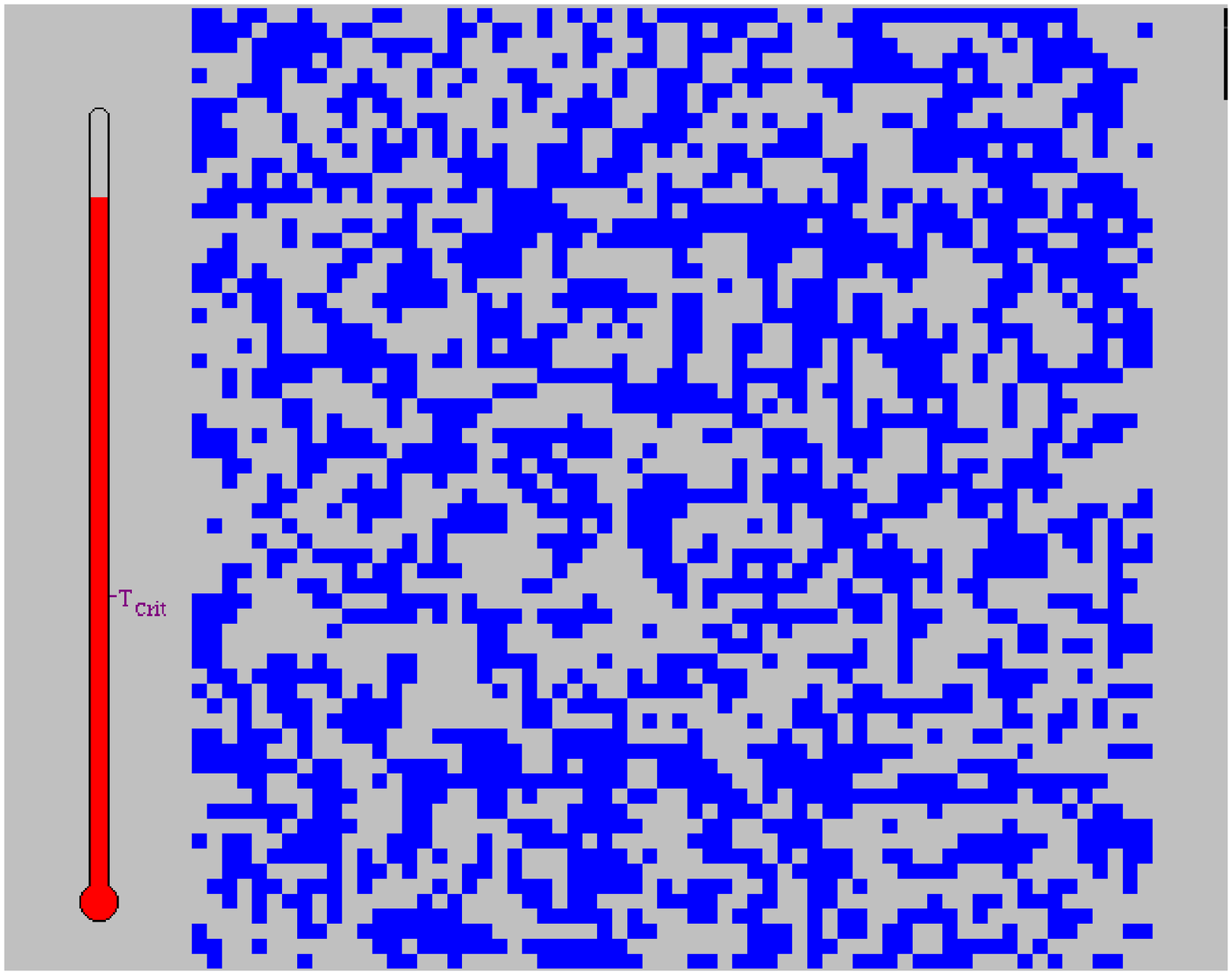} \put(-115,86){\tiny $T\!\!>\!\!T_c$}
  \end{center}
  \caption{Snapshots of simulations of the Ising model for
    subcritical case ($T<T_c$),
    critical case ($T=T_c$) and
    supercritical case ($T>T_c$)}
  \label{fig:ising}
\end{figure}

This is shown in figure \ref{fig:forfireend} where
the end configurations of different characteristic
simulations runs are given. There is indeed an
analogy to phase transitions in physical systems --
for example the Ising (see figure~\ref{fig:ising},
taken from~[\refcite{GuEvIsin98}]).

In the Ising model, for temperatures below
$T_c\approx 2.268254878$ the system is quite homogeneous
(ordered with only small fluctuations). It is quite homogeneous
as well for temperatures significantly above $T_c$ (disordered
with only small correlated regions). At $T= T_c$, however, the
picture is drastically different:
There are clusters on all length scales and the correlation
length diverges. (For all simulations on a finite
grid, the correlation length cannot diverge; thus the picture of
clusters at all length scales remains also true for $T\approx T_c$.)

Analogously we denote cases where the fire dies out within a
few timesteps and leaves behind a homogenous, almost unaffected
system as \emph{subcritical}, as illustrated in~\ref{fig:forfireend}.a.
A situation as depicted in figure~\ref{fig:forfireend}.b where one
finds burnt and unaffected clusters at all length scales is classified
as \emph{critical}. If, as depicted in figure~\ref{fig:forfireend}.c, the
fire affects the whole system and leaves behind again a homogeneous
system of burnt tree we denote this as \emph{supercritical}.

\subsection{Relation to Other Models}
\label{ssec:OtherModels}

In contrast to the majority of the models mentioned in the
introduction to section~\ref{sec:ForFirModel}, our model does
\emph{not} exhibit self-organized criticality. For example
in the Clar-Drossel-Schwabl-model~[\refcite{clar-1994}]
the combination of a growth process and repeated
ignition of forest fires is believed to push the system into
a critical state. From the exposition given in
subsection~\ref{ssec:SimRun} this is easy to
understand on an intuitive level. As long as the
system is in a subcritical state, fires will die out after
a few timesteps, resulting in net growth. In a
supercritical state, large clusters will burn, which
yields net loss of trees. The combination of both
mechanisms will lead to a critical quasi-equilibrium.

In our model no growth mechanism is present, so -- as
it is the case in typical statitical mechanics systems --
parameters have to be fine-tuned to access a critical
state. Accordingly this model provides\footnote{for
burning probability $p_b=1$, if follows the path of the
most popular models} a very close look at the
bruning process of a cluster, a ``magnifying glass''
which allows to study certain properties of  self-organized
criticality whithout the need to be very close to the
critical point. Otherwise the critical state is hard to access
numerically, since in order to deal with diverging correlation
lengths, huge grids are required to have finite-size effects
under control.

As outlined in~[\refcite{Lichtenegger:2005DA}] the
main reason why this model has been studied is the
fact that it can easily generalized to a semi-realistic
disease-spreading model. A publication on that model
is in preparation (see also comments in
section~\ref{sec:summary}).

%%%%%%%%%%%%%%%%%%%%%%%%%%%%%%%%%%%%%%%%
\section{Analysis Tools}
\label{sec:ForFirAnalysis}

After having performed the simulation (which is finished when
there are no burning trees left), it is the main task
to extract the essential information and to condense it to
a small set of significant numbers that can be easily
compared for various runs and different parameter configurations.

There are two sources of information that are easily
available: the final tree configuration and the
time series of burning trees (both depicted in figure
\ref{fig:forfireend}). The main quantities which
can be extracted from them are summarized in
table~\ref{tab:parameters}.

\subsection{Cluster Analysis}
\label{sec:for_cluster}

The methods covered in this section are in a way
related both to determining the fractal dimension of
geometric objects by box-counting and the ideas of
the renormalization group~[\refcite{Wilson:1973jj}],
in particular when formulated in the language of
block-spin transformations \`a la Kadanoff.

\begin{figure}
  \begin{center}
    \includegraphics[width=13cm]{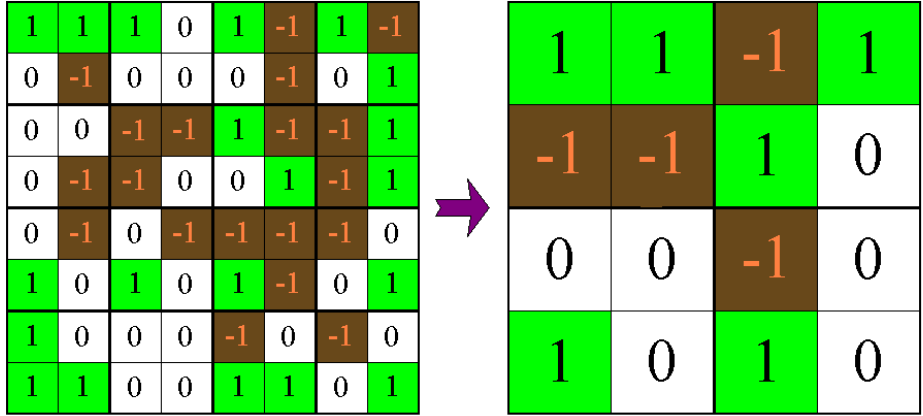}
  \end{center}
  \caption{Combining cells as described in
    equation~\eqref{eqn:cluscomp}}
  \label{fig:cluster}
\end{figure}

Starting point is a the end configuration after
a simulation run, i.e. a lattice of cell states
$S^{(0)}:=\{s_{i,\,j}\}$
(as they were defined in equation \eqref{sec:ForFirModel}:
$s=-1$ for a burnt-out, $s=0$ for an empty and $s=1$
for a living cell). Now multiple cells are combined
to new one,
\begin{equation}\label{eqn:gencompr}
  s_{i,\,j}'=f\left(s_{i_{ij,1},\,j_{ij,1}},
  s_{i_{ij,2},\,j_{ij,2}},
  s_{i_{ij,3},\,j_{ij,3}}, \ldots,
  s_{i_{ij,k},\,j_{ij,k}} \right),
\end{equation}
a procedure that yields a smaller lattice
of new cells with states $S^{(1)}=\{s_{i,\,j}'\}$. This
process can of course be repeated, resulting in a
sequence of lattice configurations:
$$\left(S^{(0)},\,S^{(1)},\,S^{(2)},\ldots,S^{(\Omega)}\right).$$
where $S^{(\Omega)}$ is a lattice consistent of too few cells
to repeat the procedure defined in equation \eqref{eqn:gencompr}.
Here, four cells were combined to a new one via
\begin{equation}
s_{i,\,j}'=\mathrm{sign}\,\left(s_{2i-1,\,2j-1}
    +s_{2i-1,\,2j}+s_{2i,\,2j-1}+s_{2i,\,2j}\right)
\label{eqn:cluscomp}
\end{equation}
a process which is illustrated in figure~\ref{fig:cluster}. Of
course, other procedures are possible as well, and others
are necessary to treat models with other types of states.

\medskip

\noindent A suitably chosen characteristic observable
$O(n)$ can be evaluated for each of these configurations,
\begin{equation}
O(n)=O(S^{(n)}),
\end{equation}
and its values may give insight to the typical length scales
present in the system. On this behalf, one can try to fit an
analytic function $O(x)$ to them and read off a the size
distribution of clusters. The most obvious choice for $O$ is
\begin{equation}
  O(n)=\frac1{N_n}\sum_{s_{i,\,j}\in S^{(n)}} \hspace{-3mm} s_{i,\,j}
\label{eqn:ffobs}\end{equation}
(with $N_n$ denoting the total number of cells in a configuration
$S^{(n)}$) which was employed here, but alternatives for $O$ are
possible as well.

Of course, after a sufficient number of recombination
steps, in most cases values of $+1$ or $-1$ will remain.
The first case corresponds to the scenario where the fire
has stopped after a comparatively small number of timesteps
(subcritical), the second is valid for mostly burnt-out
trees (supercritical). In both cases, also in view
of well-known results on fractal respectively box-counting
dimensions, $O(n)$ from equation \eqref{eqn:ffobs} can reasonably
be approximated by an exponential function:
\begin{equation}
  f(n)=1-e^{-\alpha_e n}
\label{eqn:fffite1}\end{equation}
for dominance of $+1$, respectively
\begin{equation}
  f(n)=-1+e^{-\alpha_e n}
\label{eqn:fffite2}\end{equation}
in the case of $-1$ dominating.
The constant $\alpha_e$ here is a measure of
cluster size: When $\alpha_e$ is small, this means
that there are also larger clusters of opposite
cell state; a large value of $\alpha_e$ indicates
rapid ``saturation'' and therefore the existence
only of small clusters.

In the critical case, however, living and burnt-out
trees are equally present in the whole simulation
region, so this results in values of zero repeatedly
emerging on all length scales during the combination
process, and so the exponential fit often
fails.\footnote{In the following calculations,
the fit has been done with the MATLAB \emph{nlinfit}
routine with standard parameters (version 7.1.0.183),
which reports failure if the required
precision cannot be achieved.}

So in addition, also a linear (least-square) fit has
been performed. Of course, in most cases, a linear function
is only a rather poor description -- but it is
reliable in the sense that there is a straightforward
way to find the fit and one always has a result
(more or less meaningful). So in addition to
$\alpha_e$ from equation \ref{eqn:fffite1}
respectively \ref{eqn:fffite2}, also a parameter
$\alpha_l$ for a fit
\begin{equation}
f(n)=\alpha_l\,n + C
\label{eqn:fffitlin}\end{equation}
has been retrieved. Both parameters usually
show the same qualitative behaviour. (An important
exception for $p_b\to 1$ is discussed in section
\ref{sec:forphase}.)

For the critical case the cluster size distribution is
expected to obeye a power-law; while this issue has
been left to further investigation (which has to be performed
on much larger grids and to be compared with results for
SOC models mentioned in section~\ref{sec:ForFirModel}),
a sign for an arising power-law ist the fact that in the
critical case an exponential fit~(\ref{eqn:fffite1},~\ref{eqn:fffite2})
becomes unstable and often breaks down completely.

%%%%%%%%%%%%%%%%%%%%%%%%%%%%%%%%%%%%
\subsection{Time Series Analysis}
\label{sec:for_timeser}

\begin{figure}
  \begin{center}
    \includegraphics[width=12.5cm]{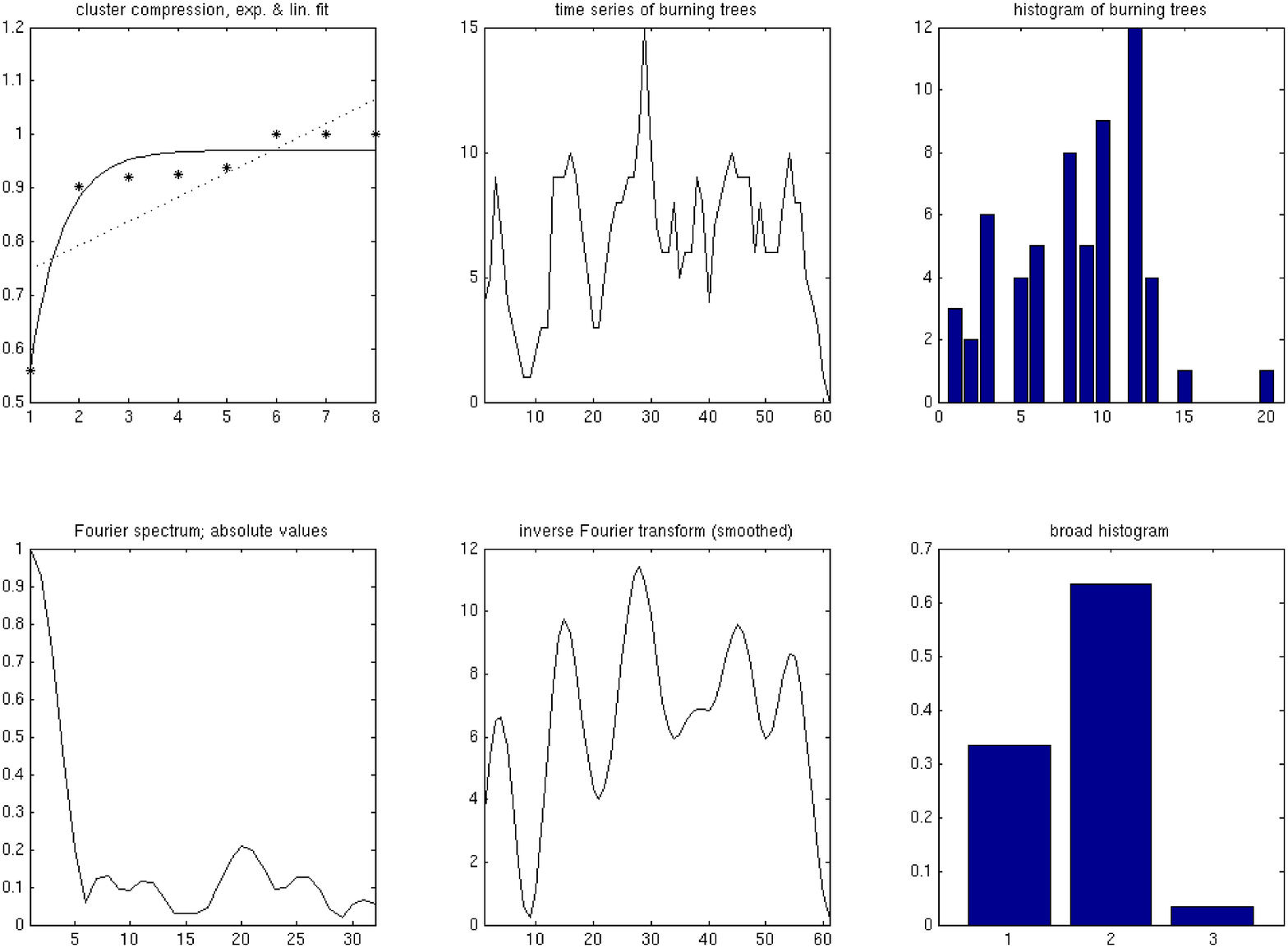}
      \put(-360,260){(a)} \\[18pt]
    \includegraphics[width=12.5cm]{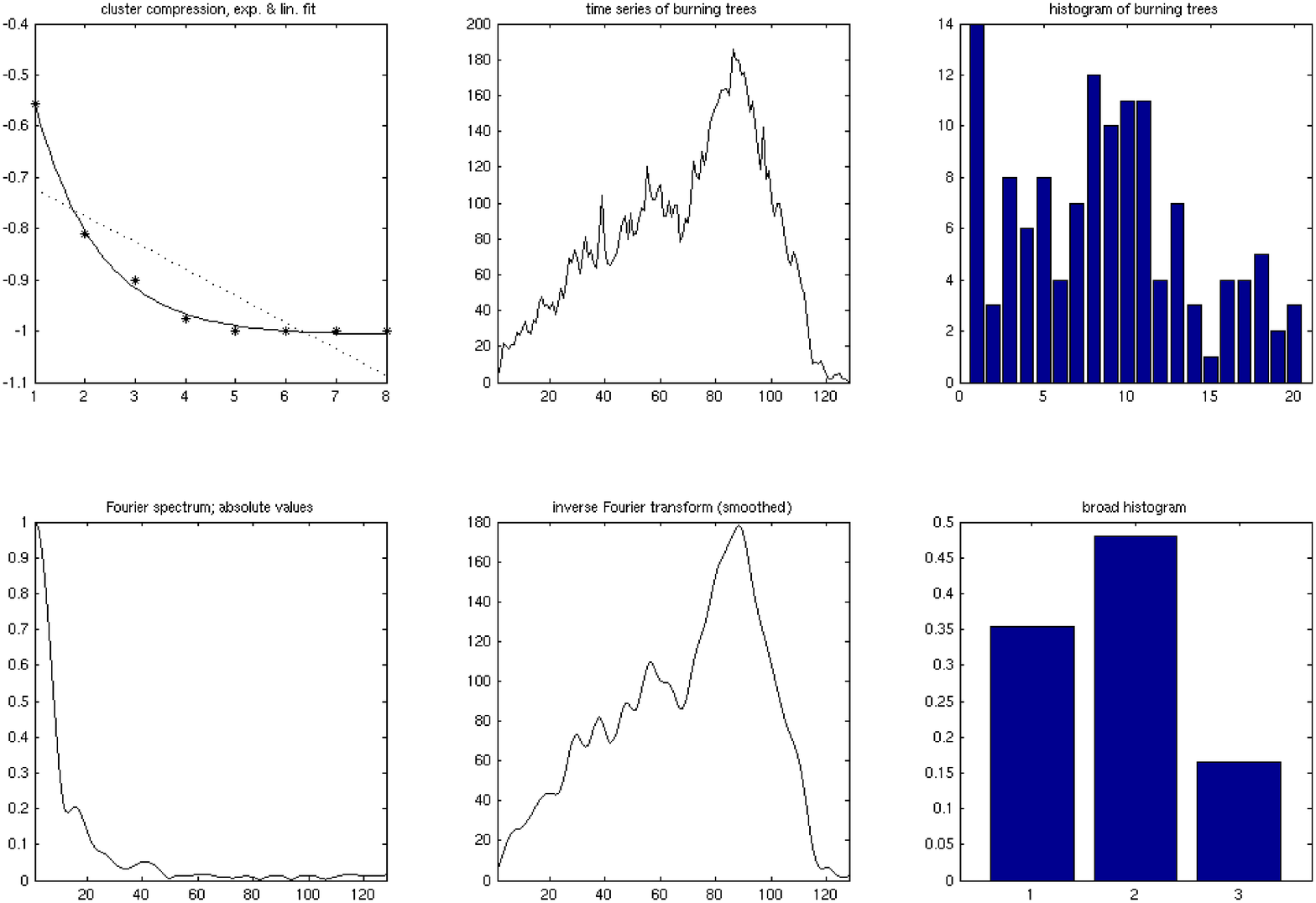}
      \put(-360,240){(b)}
  \end{center}
  \caption[Time series and cluster analysis for sub- and
        supercritical case]{Time series and cluster
        analysis for typical runs of forest fire simulations ($N=128$):
    	(a) subcritical case ($p_b=d_{\mathrm{start}}=0.61$),
    	(b) supercritical case ($p_b=d_{\mathrm{start}}=0.66$).
        In each case, the results of exponential and linear
        fit to cluster compression (equations \ref{eqn:fffite1}
        and \ref{eqn:fffite2}), the time series of burning trees,
        a corresponding histogram, the Fourier transform, a
        smoothed version of the time series and at last a very
        rough histogram are shown.}
  \label{fig:forfireanalysis}
\end{figure}

\begin{figure}
  \begin{center}
    \includegraphics[width=12.5cm]{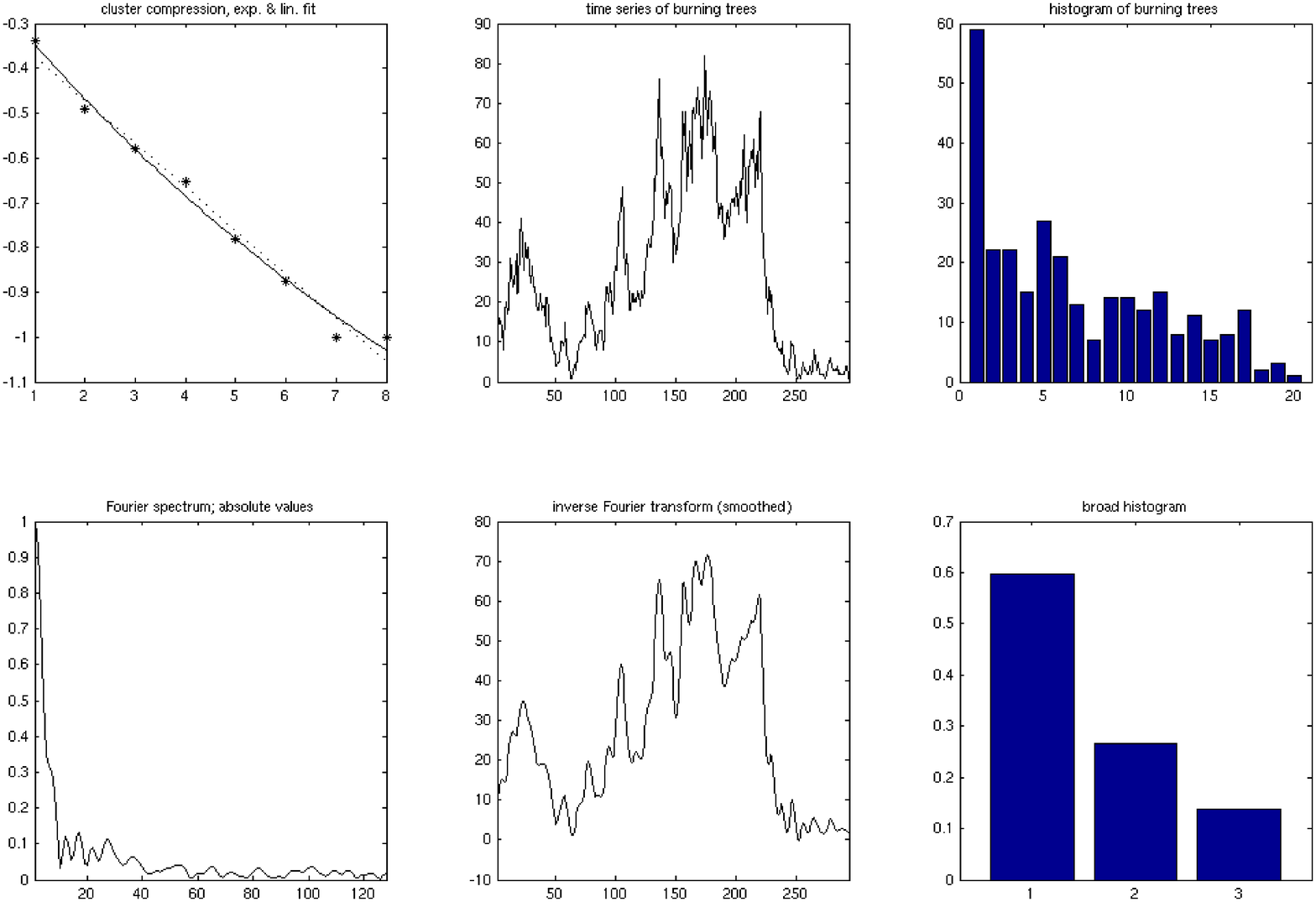}
      \put(-360,250){(a)} \\[18pt]
    \includegraphics[width=12.5cm]{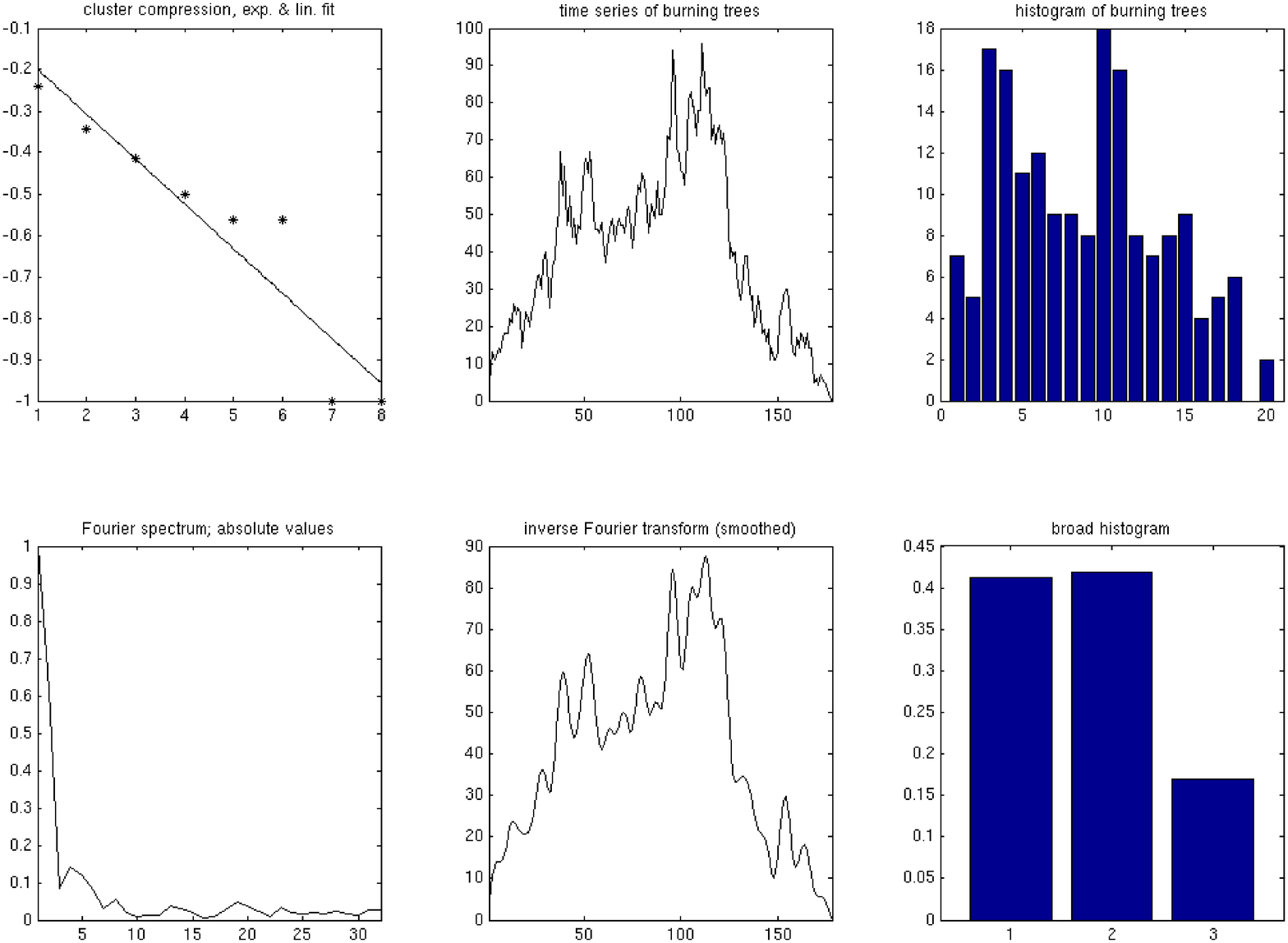}
       \put(-360,250){(b)}
  \end{center}
  \caption[Time series and cluster analysis for the critical
        case]{Time series and cluster analysis for two
        critical forest fire simulation runs
        ($N=128$, $p_b=d_{\mathrm{start}}=0.63$). As in
        figure \ref{fig:forfireanalysis}, exponential and linear
        cluster fit, the time series of burning trees, the Fourier
        transform, a smoothed time series and two histograms
        are given. In (b), however, the exponential fit
        (as a tool for cluster analysis) fails.}
  \label{fig:forfireanalysisfail}
\end{figure}

Information can also be retrieved from
the time series of burning trees. Examples for such
a time series are given in figures \ref{fig:forfirerun},
\ref{fig:forfireend}, \ref{fig:forfireanalysis} and
\ref{fig:forfireanalysisfail}. There is number of meaningful
and interesting quantities
that can be extracted from such a time series by rather
simple means. Some other parts of the analysis done,
however, rely on more sophisticated techniques, namely
discrete Fourier transform (see section \ref{sec:for_fourier}).

Some characteristic numbers can be read off
from the time series without problems: the total
burning time $t_{\mathrm{burn}}$ (timesteps from the first tree catching
fire to the last one becoming burnt-out) and the maximum number
of burning trees $N_{\mathrm{burn,max}}$.
The three main cases (subcritical, critical and supercritical)
are shown in figure \ref{fig:forfireend} (where one should mind
the different scales, both for time and number of burning trees).
The corresponding time series is in all cases quite characteristic:

\begin{itemize}
\item In the subcritical case, the fire only affects a
  small fraction of the system. In the strongly subcritical
  region, it is extinguished after a few timesteps,
  but for parameters approaching the critical region, it may last
  much longer. However, at any timestep the number of burning
  trees is comparatively small. There is of course at least one
  maximum, but there may be several of them, in any way they tend
  to be not significant.

\item In the critical case, the progression of the fire is
  far more complicated than in the other cases. Large unaffected
  clusters may form (and sometimes be partially consumed later by
  the fire). The maximum number of burning trees is smaller
  than in the supercritical case (but significantly higher
  than the subcritical values). Due to this and to the emerging
  cluster structure, the fire tends to last longer
  (up to approximately the double supercritical burning time).
 
\item In the supercritical case, all but a few trees are
  destroyed by the fire. Since the fire propagates nearly with
  maximum speed, it dies off soon after the whole system has
  been affected. There may be some small clusters, separated
  from the bulk, that allow the fire to survive some steps
  longer, but nevertheless there is usually a significant maximum
  and in virtually any case a rapid decrease (as, for example,
  in figure \ref{fig:forfireanalysis}). In the supercritical
  limit ($p_b\to 1$, $d_{\mathrm{start}}\to 1$), the time
  series approaches a sawtooth with a clear maximum\footnote{On
  an $N\times N$-grid, a simple calculation shows that there
  is a maximum of $N_{\mathrm{burn, max}}=4(N-2)$ at
  $t=\frac{N-2}2$ for an even $N$ respectively
  $N_{\mathrm{burn,max}}=4(N-1)$ at $t=\frac{N-1}2$
  for an odd grid size $N$.}, and a sudden decrease to zero
  one or two steps later.

\end{itemize}

\noindent Also the fraction of burnt trees $f_{\mathrm{burn}}$
can easily be calculated. In this special case, it is simply the
sum over the whole time series, divided by the initial number of
trees. Again, in the subcritical case,
$f_{\mathrm{burn}}$ will be small, in the supercritical it will
approach one, and for the critical case there will be a wide
range of possible values -- even for the same set of parameters.

Possibly $t_{\mathrm{burn}}$ is the best value to indicate
the phase transition from sub- to supercritical configuration.
When starting subcritical and then slowly increasing 
$p_b$ and/or $d_{\mathrm{start}}$, one will notice a
dramatic increase in $t_{\mathrm{burn}}$ when reaching
the critical region and a significant decrease when
leaving it towards supercriticality. (This is discussed
in more detail in section \ref{sec:forphase}.)

Also $N_{\mathrm{burn,max}}$ and $f_{\mathrm{burn}}$
are possible indicators for the transition, but since they
usually monotonically increase on the path from sub- to supercriticality,
it may be harder to identify the phase transition when
watching these quantities.

Another sort of information can be extracted from histograms.
Figures~\ref{fig:forfireanalysis} and~\ref{fig:forfireanalysisfail}
also include two histograms each -- one with 20 bins and one with
only three bins. So the second one contains three characteristic
numbers $h_0$, $h_1$ and $h_2$ which give the fraction
of timesteps when there were comparatively few, medium or
many burning trees.\footnote{Of course one has the relation
$h_0+h_1+h_2=1$; thus only two of these numbers are independent.}

A quantity that may become quite interesting as well,
is the number of dramatic decreases of the number of burning
trees -- but due to the rather noisy nature of the time series
this information is relatively hard to extract from the original
series. So it is postponed to the next section which deals with
Fourier analysis.

\subsection{Fourier Analysis}
\label{sec:for_fourier}

From the time series one can deduce the Fourier spectrum as
well. This has been done with the MATLAB$^{\mathrm{(R)}}$ FFT
routine (using Fast Fourier Transform). The Fourier transform is
also displayed in figures \ref{fig:forfireanalysis} and
\ref{fig:forfireanalysisfail}. In all cases, low frequencies
dominate the spectrum (which is already clear from the shape
of the time series).

In addition there tend to be several other peaks
that represent higher frequencies modulating the signal;
for very high frequencies, these peaks drown in noise.
So on the one hand, the positions and intensities
of such peaks (or at least the most important one)
may characterize some aspects of the time series.

On the other hand, by cutting off the noisy high-frequency
part of the spectrum and transforming back, one has
a smoothed version of the time series (as also shown in
figures \ref{fig:forfireanalysis} and
\ref{fig:forfireanalysisfail}). From this modified
time series, for example the number of decreases
can be read off.

These decreases have been separated in weak (less than $25\%$),
drastic ($25\%$ to $50\%$) and critical (more than $50\%$) ones,
and the counts of the latter two have been regarded as
significant for the fire spreading. Of course there will be at
least one dramatic decrease (when the fires finally dies off), which
is trivial and is therefore not counted here. Especially
in the case of critical fire propagation, there tend to be several
more such decreases. (See for example figure \ref{fig:forfirsketch}.b.)

The appearence of drastic and critical decreases seems to characterize
critical behaviour quite well, since it is more or less absent in
sub- or supercritical simulation data. This can also be understood
from the characteristics of critical fire propagation: On the edge
between affecting most cells and dying off soon, there are repeatedly
situations when only a few burning trees are left and those when
a whole cluster catches fire.
When smoothing out the time series, setting the cutoff point
usually requires some fine-tuning (and it of course may affect all
results).

For the Fourier analysis presented in section \ref{sec:ForFirResults}
the method described so far has been slightly modified: In order
to increase the comparability of results, short time series have been
expanded to a length of $2^8=256$ steps (by adding zeros after the
simulation data). So one can establish a more unified scale for the
Fourier transform.

\noindent From there, all but the lowest $32$ frequencies have been cut from
the spectrum (that contained initially $\frac12\cdot 2^8=128$ frequencies).
While this choice is of course arbitrary, it yields trustworthy results
that mostly coincide with those put forth by the best instrument for such
types of analysis -- human eye and brain.

\bigskip

\begin{figure}
\begin{center}
\begin{tabular}{|c|l|l|} \hline
$Q$ & meaning & sec. \\ \hline
$f_{\mathrm{burn}}$ & fraction of burnt trees (burning rate) & \ref{sec:for_timeser} \\
$t_{\mathrm{burn}}$ & total burning time & \ref{sec:for_timeser} \\
$N_{\mathrm{burn,max}}$ & maximum number of burning trees & \ref{sec:for_timeser} \\
$\alpha_e$ & exponential fit parameter from cluster analysis & \ref{sec:for_cluster} \\
$\alpha_l$ & linear fit parameter from cluster analysis & \ref{sec:for_cluster} \\
$h_0$ & fraction of steps with ''few'' & \\
 & ($\in [0,\,\frac{N_{\mathrm{burn,max}}}3]$)
  burning trees & \ref{sec:for_timeser}\\
$h_1$ & fraction of steps with ''several'' & \\
  & ($\in (\frac{N_{\mathrm{burn,max}}}3,\,\frac{2N_{\mathrm{burn,max}}}3]$)
  burning trees & \ref{sec:for_timeser}\\
$h_2$ & fraction of steps with ''many'' & \\
  & ($\in (\frac{2N_{\mathrm{burn,max}}}3,\,N_{\mathrm{burn,max}}]$)
  burning trees & \ref{sec:for_timeser}\\
$\nu_D$ & frequency of dominant Fourier peak & \ref{sec:for_fourier} \\
$I_D$ & intensity of peak at $\nu_D$ & \ref{sec:for_fourier} \\
$n_{\mathrm{dras}}$ & number of drastic decreases ($25\%$ to $50\%$) & \ref{sec:for_fourier} \\
$n_{\mathrm{crit}}$ & number of critical decreases ($>50\%$) & \ref{sec:for_fourier} \\ \hline
\end{tabular}
\end{center}
\caption{Summary of the quantities $Q$ eventually used as indicators of the
fire propagation (and especially the phase transition). ``sec.'' denotes the section
where the appropriate quantity is explained in detail}
\label{tab:parameters}
\end{figure}

\section{Results}
\label{sec:ForFirResults}

Here some fundamental results are presented. In contrast to the
few (but nevertheless characteristic) examples presented so far,
now the complete parameter space of the model has been scanned
systematically.

Since there are only two parameters that strongly influence system
behaviour (namely $p_b$ and $d_{\mathrm{start}}$) such a scan
is possible with moderate computational effort and the results
can be displayed in a $2D$-plot.

The grid size has little influence on any result (as long as the
lattice is sufficiently large, so that the initially burning trees
incur no significant bias and subcritical propagation affects only
a small fraction of the system). So we used a $128\times 128$ cell
grid for all simulations.

The Moore parameter $\pi_M$ has greater influence on the result,
in section \ref{sec:forinfvnf} however, we will demonstrate that
while there are quantitative changes, the qualitative behaviour
is not modified. So all other simulations have been performed
with $\pi_M=0.5$.

The parameters $p_b$ and $d_{\mathrm{start}}$ have been varied from
zero to one in steps of $\Delta p_b=\Delta d_{\mathrm{start}}=0.01$,
for each configuration there have been $40$ simulation runs where
the quantities given in section \ref{sec:ForFirAnalysis} have been
extracted.\footnote{The exponential fit parameter $\alpha_e$ has not
always been available, as it has already been discussed in
subsection~\ref{sec:for_cluster}.}

\begin{figure}
  \begin{center}
    \includegraphics[width=12.5cm]{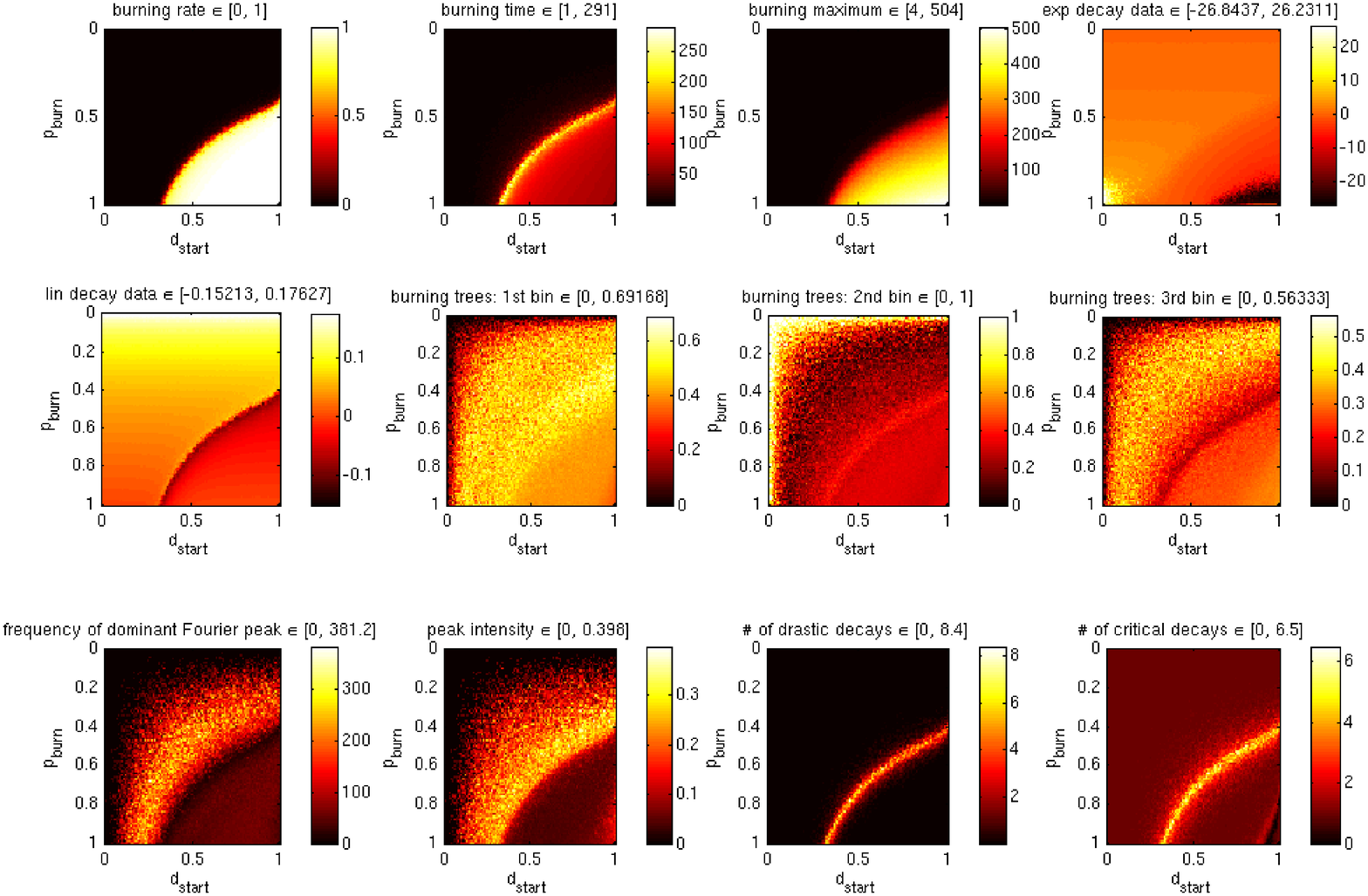}\put(-350,240){(a)} \\[18pt]
    \includegraphics[width=12.5cm]{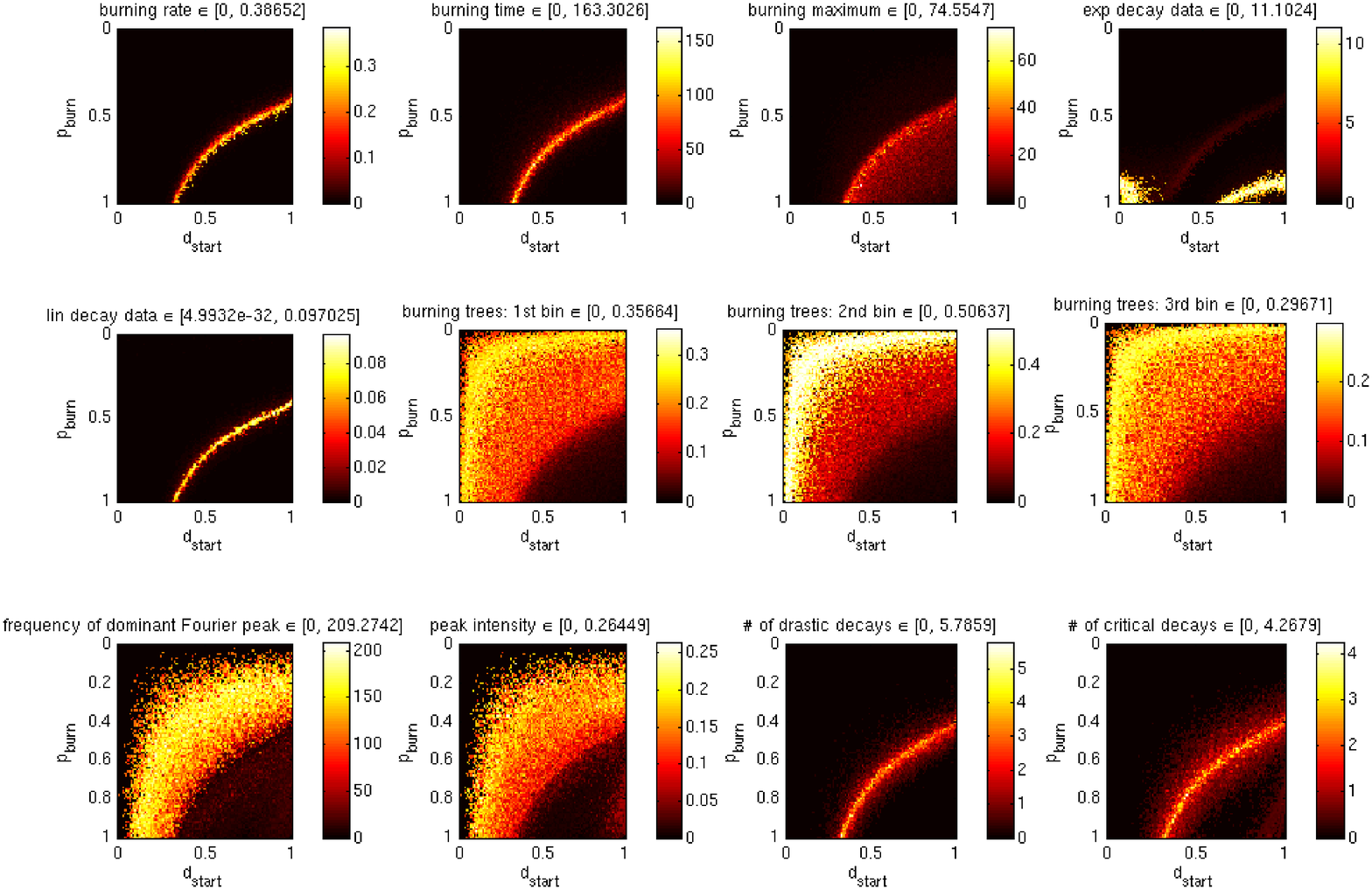}\put(-350,220){(b)}
  \end{center}
  \caption{Significant system characteristics for
     $p_b=0:.01:1$, $d_{\mathrm{start}}=0:.01:1$
     and 40 iterations:
     (a) mean values, (b) corresponding deviations. The quantities displayed
     are from left to right (i) in the first row burning rate $f_{\rm burn}$, total burning
     time $t_{\rm burn}$, maximum number of burning trees $N_{\rm burn,max}$
     and exponential decay constant $\alpha_e$, (ii) in the second row the
     linear decay constant $\alpha_l$ and the histogram values $h_1$, $h_2$, $h_3$,
     (iii) in the third row the frequence $\nu_D$ of the dominant peak, its intensity
     $I_D$, the number $n_{\rm dras}$ of drastic and $n_{\rm crit}$ of critical
     decreases.}
  \label{fig:forfirefull}
\end{figure}

Both mean values (indicated by a bar, as for example in
$\overline{t_{\mathrm{burn}}}$) and standard deviations
(here indicated by a capital delta, $\Delta$) of these quantities
are, for all configurations, given in figure \ref{fig:forfirefull}.

\subsection{The Phase Transition}
\label{sec:forphase}

As it has already been mentioned, the model exhibits a (surprisingly sharp)
''phase'' transition between two radically different kinds a behaviour,
referred to as \emph{subcritical} and \emph{supercritical}, as described in
section \ref{sec:for_timeser}. On most plots in figure \ref{fig:forfirefull}.a,
the transition line is clearly to be seen.

This line is approximately a hyperbola with
$p_b\cdot d_{\mathrm{start}}\approx C=\mathrm{const}$ where
we have the $C\approx 0.4$ for the constant.\footnote{This value
of course depends on $\pi_M$, as it can be seen in section
\ref{sec:forinfvnf}. In the present case, we have $C\approx 1-p_C$,
where $p_C\approx 0.5928$ denotes the percolation limit in
two dimensions, see~[\refcite{Aharony}].}
While the hyperbola is at least no bad approximation for the
transition line, the true form can not be deduced from the
given data -- it may even have a fractal substructure.
Its width, however, does not exceed a few percent
of change in both parameters (and probably it is still
widened due to finite size effects and the limited amount
of data given.)

The burning rate $f_{\mathrm{burn}}$ increases from below $10\%$
to above $90\%$ on this line. On the same line in configuration space,
the total burning time $t_{\mathrm{burn}}$ reaches its maximum value,
decreasing both towards the sub- and the supercritical region. The burning
maximum $\overline{N_{\mathrm{burn,max}}}$
begins to increase from significantly below $100$ to more than $200$,
but this transition is not that sharp and there is a further increase
up to about $\overline{N_{\mathrm{burn,max}}}=500$
for $p_b\to 1$ and $d_{\mathrm{start}}\to 1$.

For the exponential decay parameter $\alpha_e$ there seems to be no clear
transition at all, this impression, however, is created only by the plot scales.
Since there are both regions with $\overline{\alpha_e}>25$ (for $d_{\mathrm{start}}\ll 1$
and $(1-p_b)\ll 1$) and $\overline{\alpha_e}<-25$ (for $(1-d_{\mathrm{start}})\ll 1$
and $(1-p_b)\ll 1$), the line of $\overline{\alpha_e}\approx 0$ that indicates the
transition is hardly visible.

Since the range of values is more limited for $\alpha_l$, the phase transition
is easier to recognize for this fit parameter. The transition again has
$\overline{\alpha_l}\approx 0$, but now, such values are also reached for
$p_b\to 1$, in region where $\overline{\alpha_e}$ takes its extreme values.
This, however, is clear from the nature of the different fits. In these
regions, the exponential decay is very fast, so already from the second step on,
the data points lie approximately on a straight horizontal line, and therefore
we have $\alpha_l\approx 0$.

The histogram data ($h_0$ to $h_2$) is more difficult to interpret. Also in
these plots, the transition line is visible, but it is only one of several
regions where interesting changes occur. Also for frequency and intensity of
the Fourier spectrum, the interpretation is not that easy. For the product
$p_b\cdot d_{\mathrm{start}}$ being small (i.e. $\ll 1$), one has a low
maximum intensity at a low frequency. This can be accounted to a very short
time series, yielding only a few frequencies. (The following zeros affect
the scaling, but not the frequencies present in the signal.)

With $p_b$ and/or $d_{\mathrm{start}}$ increasing, first frequency and intensity
of the peak tend to increase, indicating a signal fluctuating an a small
scale. Near the critical line, however, there is a significant decrease in
mean peak frequency, indicating modulation of the signal on much larger
scale. This is also consistent with the increase of the number of drastic
or critical decreases -- one often has several times when the fire has nearly
died off, but also periods of chain-reaction-like propagation.

\medskip

\noindent These results (especially when interpreting the model as one of
disease propagation) are in a way unsettling: Even an small change of either
population density or susceptibility may introduce the transition
from an isolated (endemic) disease to an epidemic or even pandemic one.
Especially a decrease in resistance can easily be induced by natural
catastrophes or starvation -- the black death in Europe from 1347 to
1353 or the Spanish influenza after world war I are likely to be interpreted
this way.

On the other hand, they also indicate that even small changes (in
population density as well as in susceptibility) can reduce an
epidemic diseases to a mere nuisance. However, one should be
cautious when transferring the results of such models to systems
which they are not intended to be described (as it is human society).

\subsection{Influence of the Moore parameter $\pi_M$}
\label{sec:forinfvnf}

As defined in \ref{sec:defneigh}, the Moore parameter $\pi_M$ is a way
to formally mediate the transition from a von Neumann to a
Moore neighbourhood. In the previous simulations, $\pi_M=0.5$
has been employed. Now we study the effects of changing $\pi_M$.

\begin{figure}
  \begin{center}
    \includegraphics[width=12.5cm]{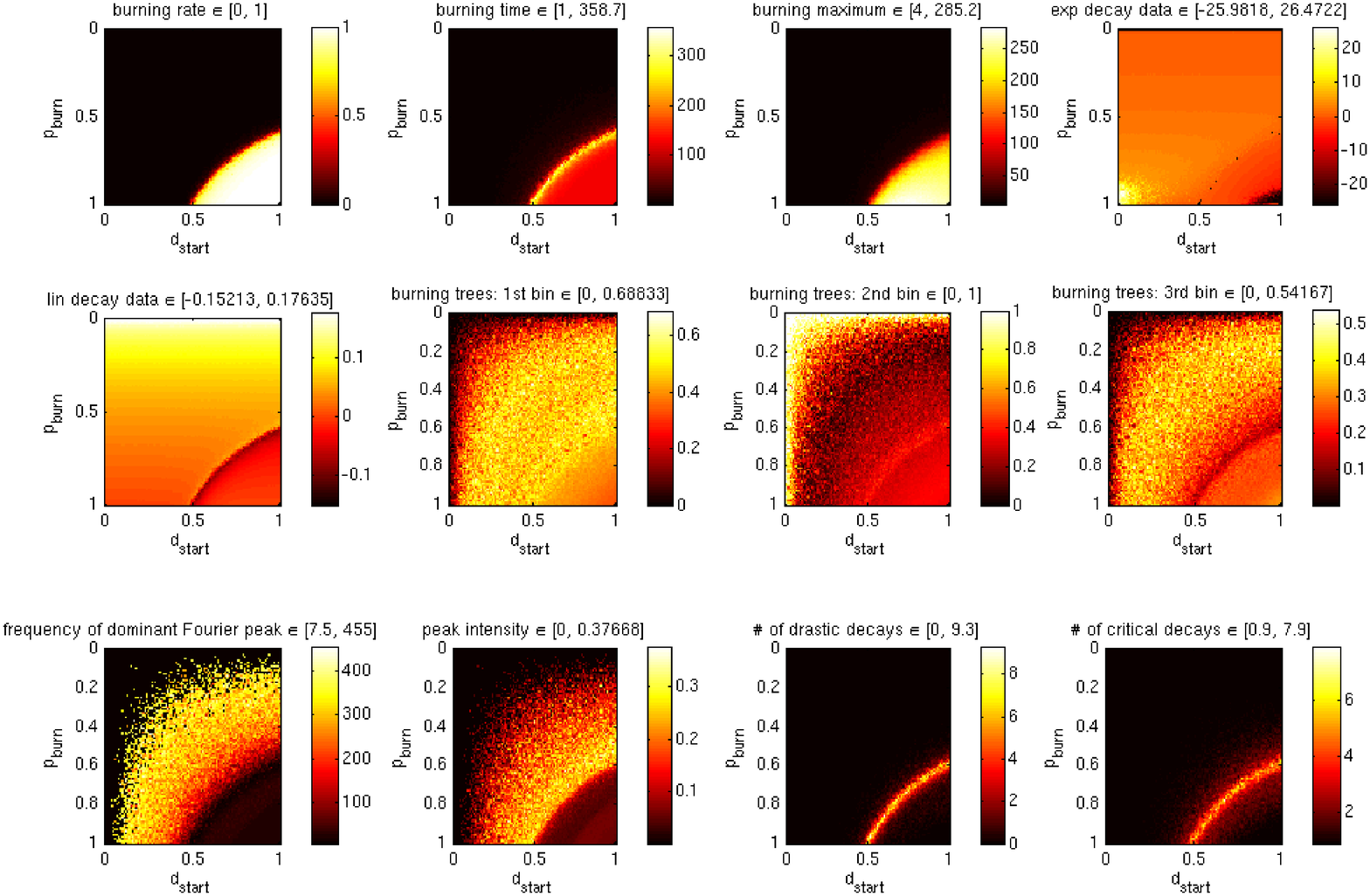} \put(-350,240){(a)} \\[18pt]
    \includegraphics[width=12.5cm]{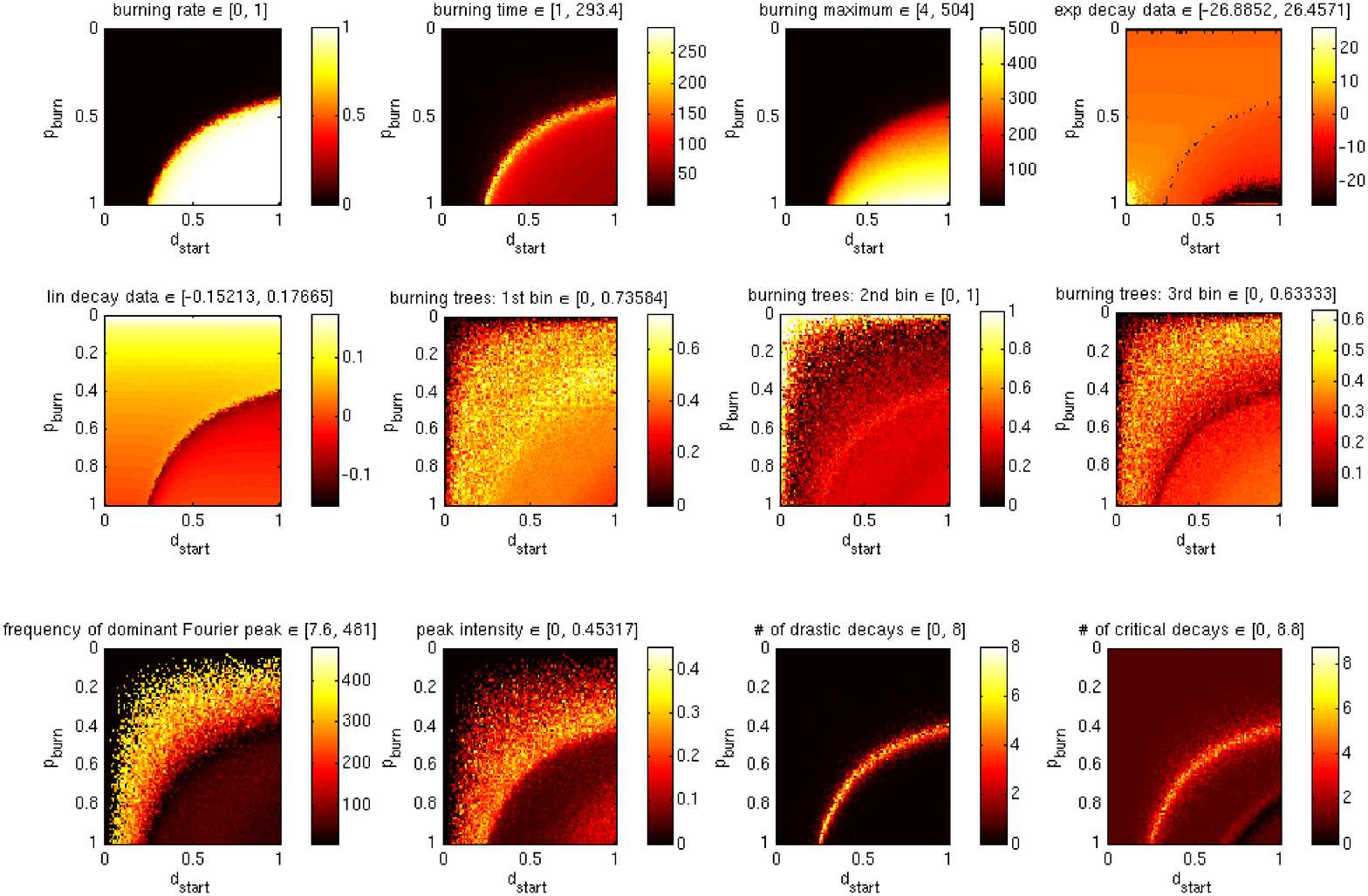} \put(-350,240){(b)}
  \end{center}
  \caption{Results of changing the \emph{Moore parameter}:
     (a) $\pi_M=0$, (b) $\pi_M=1$;
     ($p_b=0:.01:1$, $d_{\mathrm{start}}=0:.01:1$, 
     10 iterations each). The quantities displayed are
     the same as in figure~\ref{fig:forfirefull} except that
     only mean values, not deviations are shown.}
  \label{fig:forfirevnf}
\end{figure}

Therefore simulations as described in \ref{sec:forphase} have been performed
for $\pi_M=0$ (pure von Neumann neighbourhood) and $\pi_M=1$
(full Moore neighbourhood). The results are given in figure
\ref{fig:forfirevnf}.

As it is also the case in figure \ref{fig:forfirefull}, there is a clear
phase transition (indicated especially by $f_{\mathrm{burn}}$, $t_{\mathrm{burn}}$ and
$\alpha_l$), but the transition line is shifted -- towards a larger
product $p_b\cdot d_{\mathrm{start}}$ for $\pi_M=0$, towards a smaller
$p_b\cdot d_{\mathrm{start}}$ for $\pi_M=1$. (This result is fully
consistent with what could be expected, since a higher $\pi_M$ results
in a higher fire propagation probability.)

Since the behaviour is qualitatively the same for all
choices of $\pi_M$, it seems justified to perform this type
of simulation with just one value for the Moore parameter
(where we have used $\pi_M=0.5$).

\newpage

\subsection{Individual Susceptibilities}
\label{sec:ForFirAdv}

While the model presented here is of course limited in its
range of applications, there is nevertheless a simple modification
that can be implemented with virtually no additional
effort -- the substitution of a global burning probability
by an individually varying one.

\begin{figure}
  \begin{center}
    \includegraphics[width=12.5cm]{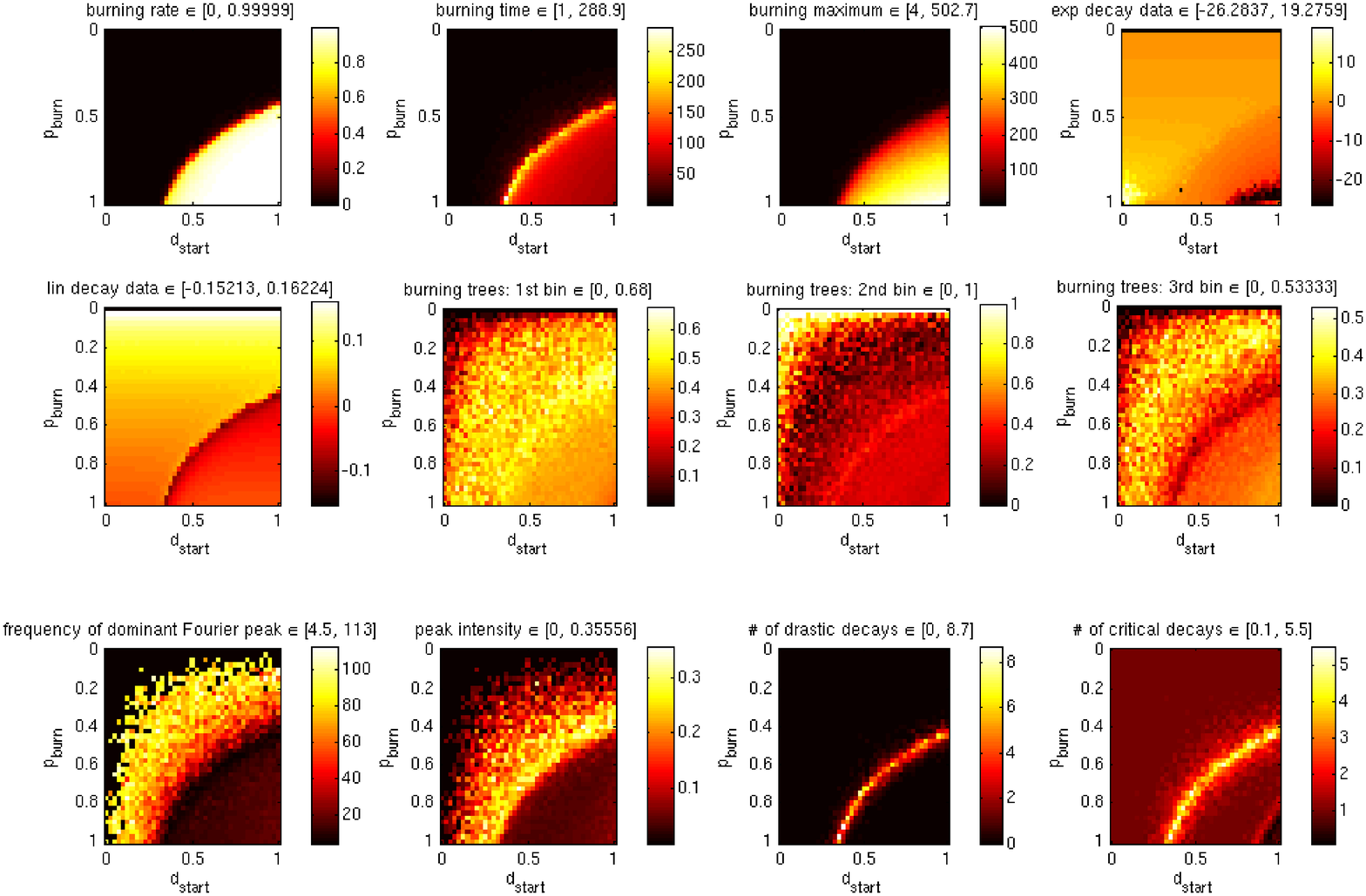} \put(-350,230){(a)} \\[18pt]
    \includegraphics[width=12.5cm]{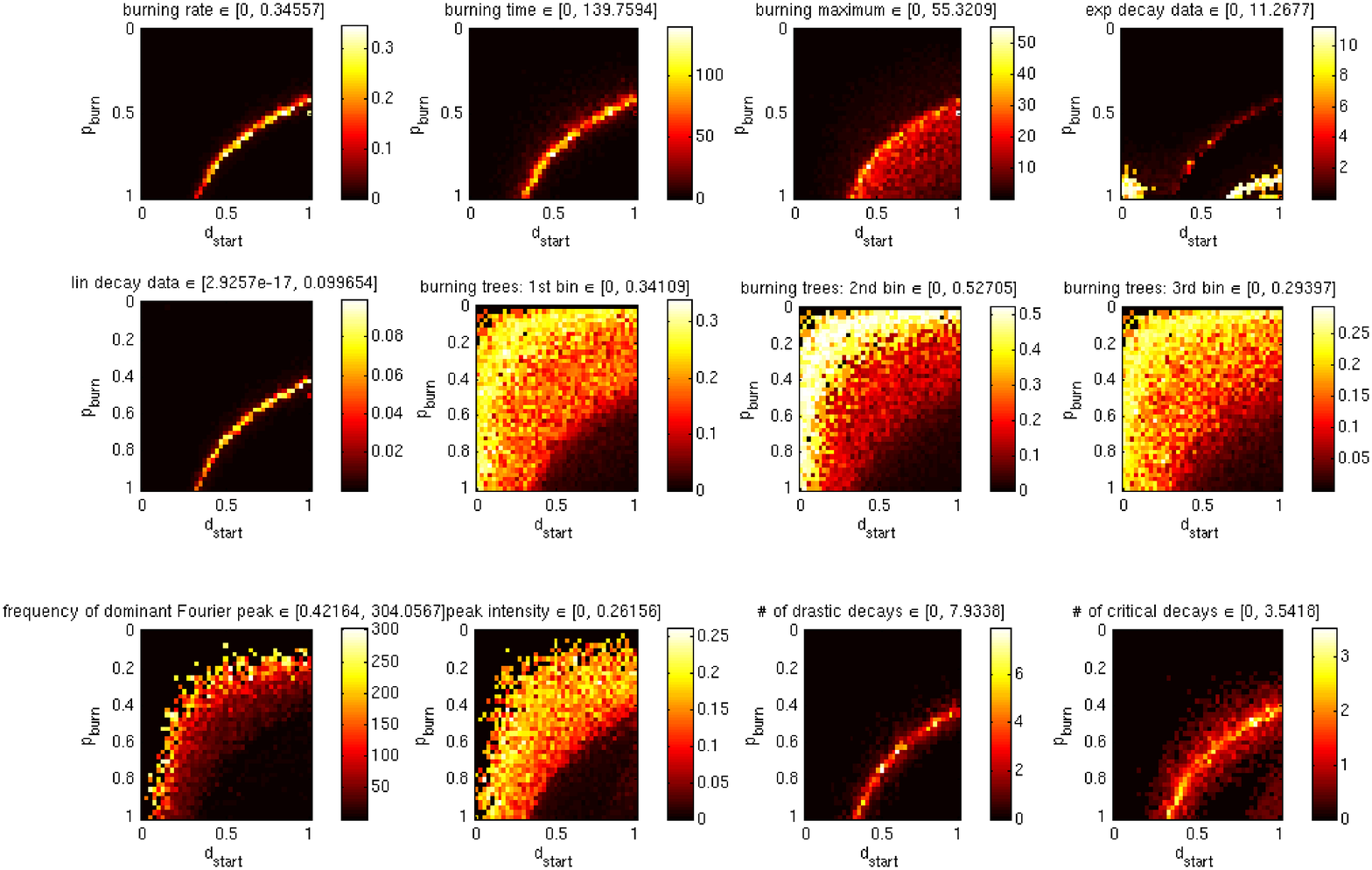} \put(-350,230){(b)}
  \end{center}
  \caption[Influence of $\sigma_b=0.1$]
      {Influence of $\sigma_b=0.1$,
      simulations on a $128\times 128$-grid with $\pi_M=0.5$,
      $\Delta p_b=\Delta d_{\mathrm{start}}=0.025$
      and 10 iterations: (a) data, (b) deviations.
      The quantities displayed are
      the same as in figure~\ref{fig:forfirefull}}
  \label{fig:forsigma1}
\end{figure}

\begin{figure}
  \begin{center}
    \includegraphics[width=12.5cm]{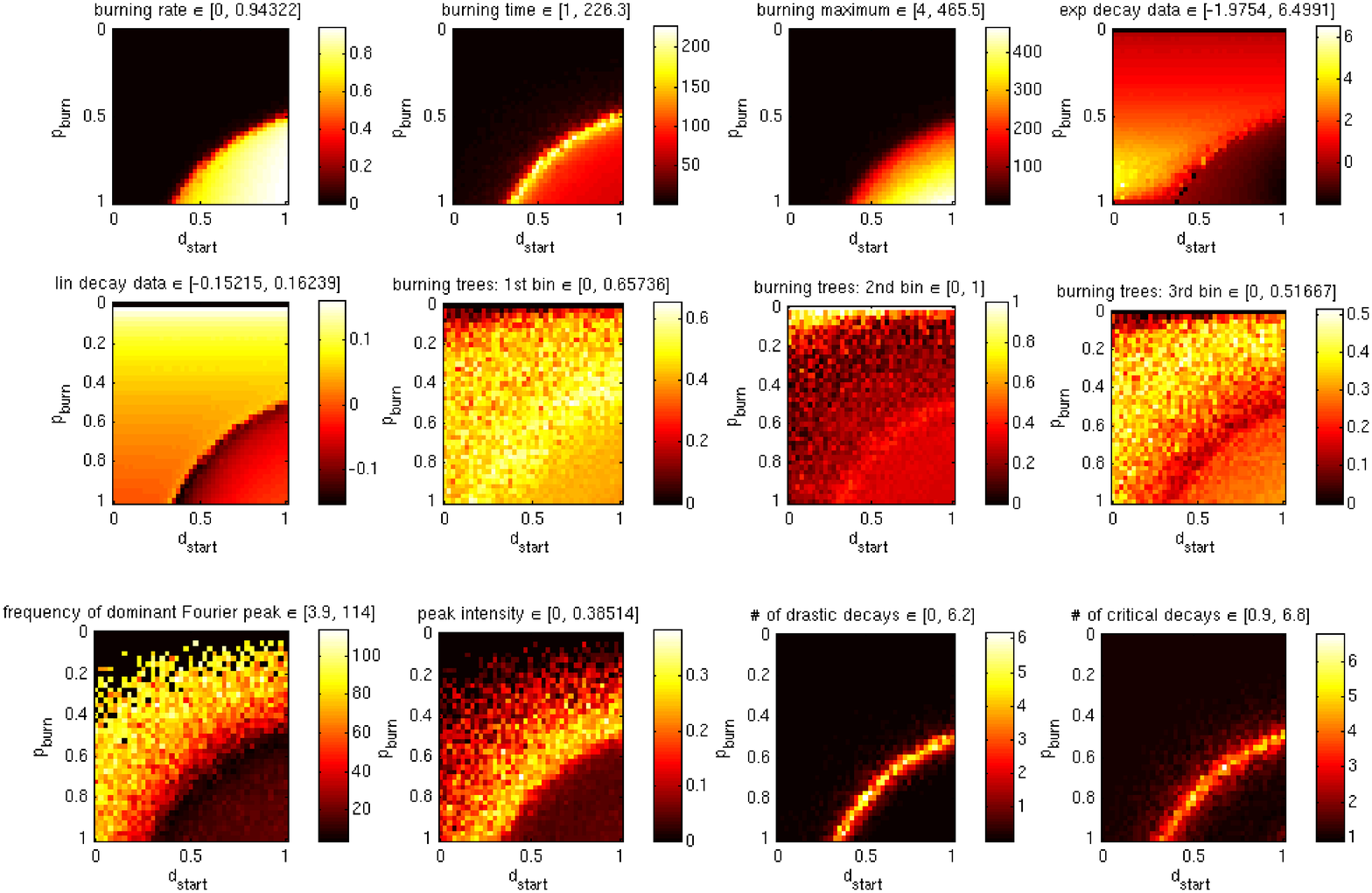} \put(-350,230){(a)} \\[18pt]
    \includegraphics[width=12.5cm]{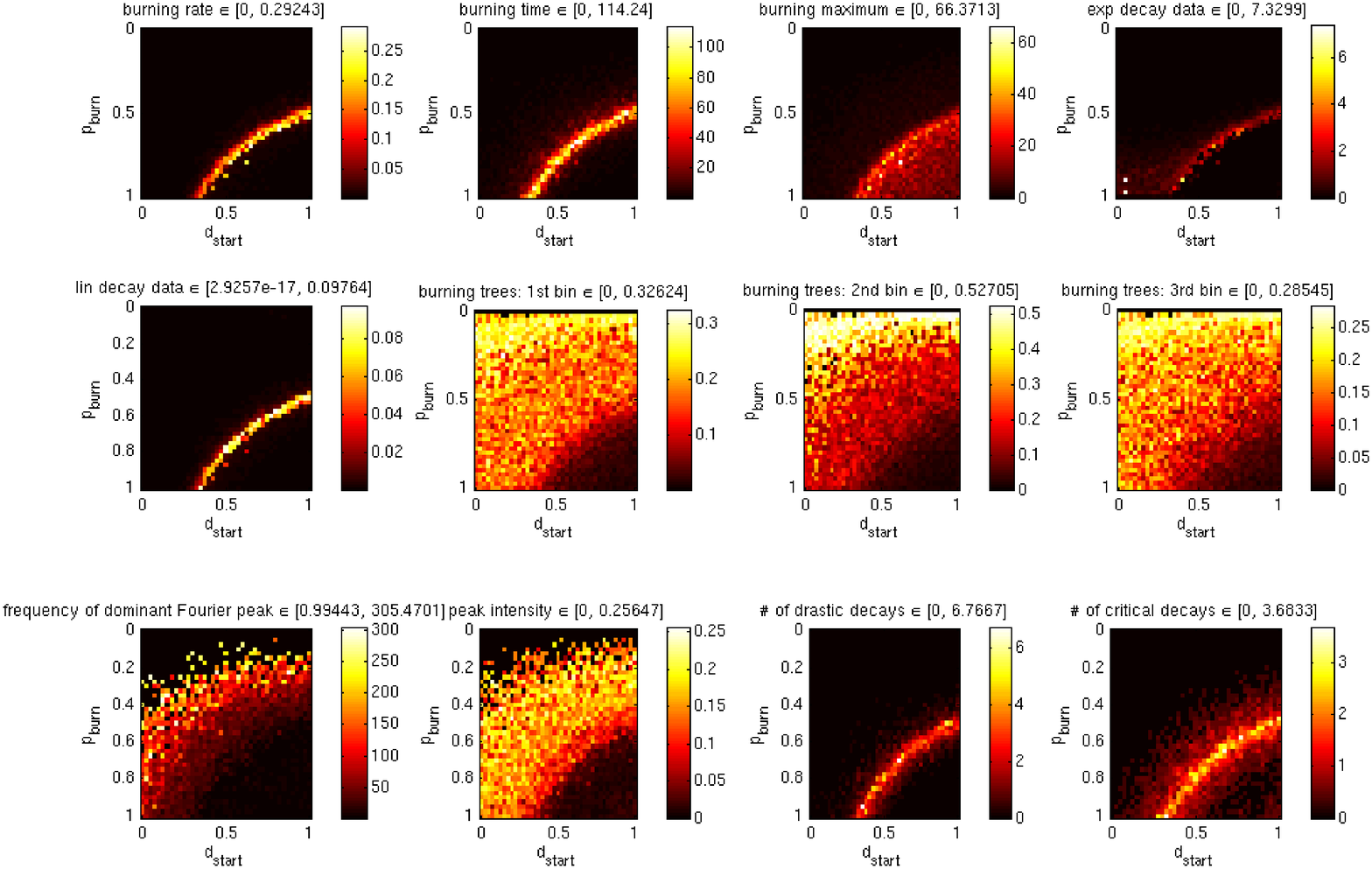} \put(-350,230){(b)}
  \end{center}
  \caption[Influence of $\sigma_b=0.5$]
      {Influence of $\sigma_b=0.5$,
      simulations on a $128\times 128$-grid with $\pi_M=0.5$,
      $\Delta p_b=\Delta d_{\mathrm{start}}=0.025$
      and 10 iterations: (a) data, (b) deviations. The
      quantities displayed are
      the same as in figure~\ref{fig:forfirefull}}
  \label{fig:forsigma3}
\end{figure}

So we define a matrix $\mathbf{P}$
with elements $p_{i,j}$ that describe the individual
burning probabilities of the cells $(i,\,j)$ and
replace equation~\eqref{eqn:forfirp12} by
\begin{equation}\label{eqn:forfirp12extended}
  p_{[1\to2]}\left(i,\,j,\, t\,|\;\mathcal{N}_{t-1}\right) =
  \delta_{s_{i,j},1} \cdot \left\{ 1-(1-p_{i,j})^{N_{\mathrm{burn}, t-1}} \right\}
\end{equation}
The distribution of the elements $p_{i,j}$ can in principle
be chosen arbitrarily, but it seems reasonable to pick them from
a Gaussian distribution centered at $p_b$:
\begin{equation}
  p_{i,j}=\mathrm{med}\left\{0,\,G(p_b;\,\sigma_b),\,1\right\}
    =\mathrm{med}\left\{0,\,p_b+\sigma_b\,G(0;\,1),\,1\right\}
\label{eqn:forfirpgauss}\end{equation}
with $\mathrm{med}$ denoting the median value
and $G(\mu;\,\sigma)$ a random variable from a Gaussian
distribution with mean $\mu$ and variance $\sigma$.
For $\sigma_b=0$ this is exactly the model already studied, but
for $\sigma_b\ne 0$ (where we use without loss of generality
$\sigma_b> 0$) new effects may arise. So we studied the various cases;
the results for a distribution ($\sigma_b=0.1$) and a broad distribution
($\sigma_b=0.5$) are displayed in figure \ref{fig:forsigma1},
and figure \ref{fig:forsigma3}. (Compared to figures \ref{fig:forfirefull}
and \ref{fig:forfirevnf}, the resolution of configuration space
sampling has been reduced due to the computational cost.)

Qualitatively, there are no severe changes, although the
phase transition line widens for higher values of $\sigma_b$.
This seems clear, since a wide range of burning probabilities
allows some sub- and supercritical configurations to obtain
characteristics from the critical case. On the other hand,
critical configurations may gain more characteristics
of either sub- or supercriticality as well, so the transition
is smeared out over a larger region of parameter space.

The magnitude of the standard deviation does not
change significantly compared to simulations performed for
$\sigma_b=0$ (but with otherwise identical parameters). This may
be somehow surprising, since one could expect the deviation in
$p_{\mathrm{burn}}$ to result in larger differences between
simulation runs with the same choice of parameters.

But of course for the critical or nearly critical case, even
with $\sigma_b=0$ the same set of parameters can lead to
radically different outcomes. So the deviations can be
expected to be large already in this case; any change
of $\sigma_b$ will not significantly increase them.
Nevertheless there are several significant quantitative changes
because of the modification of the width $\sigma_b$:

\begin{itemize}
\item The burning rates $f_{\mathrm{burn}}$ are reduced in the supercritical
  region, which is clear since also in the supercritical case now there will
  remain several trees with a burning probability considerably lower than
  average.
\item The total burning time $t_{\mathrm{burn}}$ is reduced as well as
  the burning maximum $N_{\mathrm{burn,max}}$, probably due to similar
  reasons.
\item While the change in the linear fit parameter $\alpha_l$ is not significant,
  this is not true for the exponential fit via $\alpha_e$. Large values of
  $|\alpha_e|$ indicate a rapid saturation in the cell combination analysis
  -- and such a saturation will definitely be slowed down by regions of
  considerably increased or reduced burning probability. So $|\alpha_e|$
  is considerably smaller for larger values of $\sigma_b$.
\item The dominant Fourier peak is shifted towards lower frequencies
  while the intensities remain similar.
\end{itemize}

\section{Summary and Outlook}
\label{sec:summary}

We have studied an enhanced forest fire
model that shows a surprisingly sharp phase transition
in respect to the two main parameters, namely
tree density $d_\mathrm{start}$ and burning probability
$p_b$ for a healthy tree with a burning
neighbour.

In a large region of configuration space, the
fire (or pathogen) vanishes within a few timesteps,
whereas in most of the remaining space, the fire
affects almost the whole population. The
border between these two outcomes is a small region
of ``critical'' behaviour that can be characterized
by cluster formation (on any length scale)
and exceptional values for many characteristic
quantities. (For example the total burning time
tends to be more than twice as long as in any
other case.)

A redefinition of neighbourhood, controlled by the
\emph{Moore-parameter} $\pi_M$ did not change the
qualitative behaviour, but influenced the position
of the line of critical behaviour in phase space.
In contrast, the transition to individual susceptibility
had only small influence on the global characteristics
of the system also on quantitative level.

The model can easily be re-interpreted as a disease
spreading model, where susceptible, infected and
removed individuals replace living, burning and burnt-out
trees. Our results thus may indicate that the transition
from an endemic disease (which can only affect a small
fraction of the population) to an epidemic (or,
depending on the system, even a pandemic) can be
driven by a change of just a few percent in one
crucial parameter.

While some analysis is to be done also for this model
(like the check whether for the critical case a power law
arises for the cluster size distribution), there are several
possible extensions which remain to be investigated.
Allowing empty cells to be (re)populated changes the
model significantly and reduces the importance of the
initial density. Some results can be found
in~[\refcite{Lichtenegger:2005DA}]; a comprehensive
article on this topic is in preparation.

\end{document}